\documentclass[onecolumn]{emulateapj}
\usepackage{placeins}
\shorttitle{Spectrum of $\kappa$ Andromeda b}
\shortauthors{Wilcomb et al.}

\begin{document}

\title{Moderate-Resolution $K$-Band Spectroscopy of Substellar Companion $\kappa$ Andromedae b}
\author{Kielan K. Wilcomb\altaffilmark{1}, Quinn M. Konopacky\altaffilmark{1}, Travis S. Barman\altaffilmark{2}, Christopher A. Theissen\altaffilmark{1,6}, Jean-Baptiste Ruffio\altaffilmark{3,4}, Laci Brock\altaffilmark{2}, Bruce Macintosh\altaffilmark{3}, Christian Marois\altaffilmark{5}}
\altaffiltext{1}{Center for Astrophysics and Space Sciences,
  University of California, San Diego, La
  Jolla, CA 92093, USA; kwilcomb@ucsd.edu}
\altaffiltext{2}{Lunar and Planetary Laboratory, University of Arizona, Tucson, AZ 85721, USA}
\altaffiltext{3}{Kavli Institute for Particle Astrophysics and Cosmology, Stanford University, Stanford, CA 94305, USA}
\altaffiltext{4}{Department of Astronomy, California Institute of Technology, Pasadena, CA 91125, USA}
\altaffiltext{5}{NRC Herzberg Astronomy and Astrophysics, 5071 West Saanich Rd, Victoria, BC V9E 2E7, Canada}
\altaffiltext{6}{NASA Sagan Fellow}


\begin{abstract}
    We present moderate-resolution ($R\sim4000$) $K$ band spectra of the “super-Jupiter,” $\kappa$ Andromedae b.  The data were taken with the OSIRIS integral field spectrograph at Keck Observatory. The spectra reveal resolved molecular lines from H$_{2}$O and CO. The spectra are compared to a custom $PHOENIX$ atmosphere model grid appropriate for young planetary-mass objects. We fit the data using a Markov Chain Monte Carlo forward modeling method. Using a combination of our moderate-resolution spectrum and low-resolution, broadband data from the literature, we derive an effective temperature of $T_\mathrm{eff}$ = 1950 - 2150 K, a surface gravity of $\log g=3.5 - 4.5$, and a metallicity of [M/H] = $-0.2 - 0.0$.  These values are consistent with previous estimates from atmospheric modeling and the currently favored young age of the system ($<$50 Myr). We derive a C/O ratio of 0.70$_{-0.24}^{+0.09}$ for the source, broadly consistent with the solar C/O ratio.  This, coupled with the slightly subsolar metallicity, implies a composition consistent with that of the host star, and is suggestive of formation by a rapid process. The subsolar metallicity of $\kappa$ Andromedae b is also consistent with predictions of formation via gravitational instability. Further constraints on formation of the companion will require measurement of the C/O ratio of $\kappa$ Andromedae A. We also measure the radial velocity of $\kappa$ Andromedae b for the first time, with a value of $-1.4\pm0.9\,\mathrm{km}\,\mathrm{s}^{-1}$  relative to the host star. We find that the derived radial velocity is consistent with the estimated high eccentricity of $\kappa$ Andromedae b.

\end{abstract}

\section{Introduction}

The new era of direct imaging of exoplanets has revealed a population of Jupiter-like objects that orbit their host stars at large separations ($\sim$10--100 AU; \citealt{Bowler16,Nielsen19,vigan20}).  These giant planets, with masses between $\sim$2--14 $M_\mathrm{Jup}$ and effective temperatures between $\sim$500--2000 K, are young ($\sim$15--200 Myr) compared to exoplanets discovered through other methods (e.g., Doppler spectroscopy, transit, gravitational microlensing) because their detectability is enhanced at young ages (e.g., \citealt{Baraffe08}). The formation of these gas giant planets has traditionally been challenging for the two main planet formation models, core (or pebble) accretion and gravitational instability (e.g., \citealt{Dodson-Robinson09}).

Some planet formation scenarios influence a planet’s final atmospheric composition more than others. A potential connection between formation and composition highlights the importance of studying the properties of exoplanet atmospheres.  It has long been suggested that the compositions of giant planets in our Solar System were likely determined by their initial location in the protoplanetary disk and the accretion they experienced (e.g., \citealt{Owen99}).  For example, the ratio of the abundances of carbon and oxygen (C/O) in a Jovian planet atmosphere has been suggested as a potential way to trace the location and mechanism of formation (e.g., \citealt{Madhu11}).  To estimate elemental abundances, however, we need a detailed understanding of chemical and dynamical histories of the giant planets’ atmospheres.  The luminosity and effective temperature of a giant planet decreases with time causing its atmosphere to undergo considerable changes even over a short period of time equal to the age of the directly imaged planet population, and certainly over a few billion years. In particular, the vertical mixing timescales will change as the planet’s atmospheric dynamics evolve and as the radiative-convective boundary moves to higher pressures.  Changes in temperature and pressure will also result in changes in the atmospheric abundances of gasses and condensates. The composition of their atmospheres could be further altered by continued accretion of solid bodies from the planetary disk, or mixing inside the metal-rich core (e.g., \citealt{Mousis09}). Important trace molecules (H$_{2}$O, CH$_{4}$, CO$_{2}$, CO, NH$_{3}$, and N$_{2}$) of giant planets are greatly impacted by these complex chemical and physical processes that occur over time (e.g., \citealt{Zahnle14}). 

Because of these challenges, detailed abundance measurements for certain species, such as oxygen, have been challenging for the planets in our Solar System.  For Saturn, only upper limits on the C/O ratio have been measured \citep{wong04,visscher05}. For Jupiter, previous estimates of C/O were impacted by inconclusive findings on the water abundance in the atmosphere from the Galileo probe.  Using Juno data, \cite{li20} recently measured the water abundance in the equatorial zone as 2.5$^{+2.2}_{-1.6}$ \(\times 10^3\) ppm, suggesting an oxygen abundance roughly three times the Solar value.  The directly imaged planets offer an interesting laboratory for pursuing detailed chemical abundances, as they have not undergone as many complex changes in composition as their older counterparts.

The $\kappa$ Andromedae ($\kappa$ And) system consists of a B9V-type host star with a mass of $\sim$2.7 $M_{\odot}$ and a bound companion, $\kappa$ And b \citep{Carson13}. This system is one of the most massive stars known to host an extrasolar planet or low-mass brown dwarf companion. $\kappa$ And b has been described as a “super-Jupiter,” with a lower mass limit near or just below the deuterium burning limit \citep{Carson13,Hinkley13}.

\cite{Zuckerman11} proposed that $\kappa$ And is a member of the Columba association with an age of $\sim$30 Myr, leading \cite{Carson13} to adopt that age and estimate $\kappa$ And b to have a mass $\sim$12.8 $M_\mathrm{Jup}$ with DUSTY evolutionary models \citep{Chabrier00}. However, \cite{Hinkley13} suggested that $\kappa$ And b had a much older isochronal age of 220 ${\pm}$ 100 Myr, a higher surface gravity ($\log g \approx 4.33$ as opposed to $\log g \sim 4$ for 30 Myr), and a mass of 50$^{+16}_{-13}$ $M_\mathrm{Jup}$ by comparing its low-resolution $YJH$-band spectra with empirical spectra of brown dwarfs.  \cite{Bonnefoy14} derived a similar age to \cite{Carson13} of 30$^{+120}_{-10}$ Myr based on the age of the Columba association and a lower mass limit of 10 $M_\mathrm{Jup}$ based on “warm-start” evolutionary models, but did not constrain the surface gravity. More recent studies of $\kappa$ And b by \cite{currie18} and \cite{uyama20} have concluded the object is low gravity ($\log g \sim 4$--4.5) and resembles an L0--L1 dwarf. Other studies focusing on the host star found the system to be young ($t$ $\sim$ 30--40 Myr; \citealt{david15,brandt15}). Using CHARA interferometry, \cite{jones16} constrained the rotation rate, gravity, luminosity, and surface temperature of $\kappa$ And A and compared these properties to stellar evolution models, showing that the models favor a young age, 47$^{+27}_{-40}$ Myr, which agrees with a more recent age estimate of $42^{+6}_{-4}$ Myr for the Columba association by \cite{bell15}. 

Understanding the orbital dynamics of exoplanets can also put constraints on formation pathways.  Radial velocity measurements can be used to break the degeneracy in the orientation of the planets' orbital plane. While astrometric measurements from imaging are ever increasing in precision (e.g., \citealt{wang2018}), measuring the radial velocity (RV) of directly imaged exoplanets is challenging due to the required higher spectral resolution balanced with their faintness and contrast with respect to their host stars. The first RV measurement of a directly imaged planet was $\beta$ Pictoris b using the Cryogenic High-Resolution Infrared Echelle Spectrograph (CRIRES, R=100,000) at the Very Large Telescope (VLT; \citealt{kaeufl04}). An RV of -15.4 $\pm$ 1.7 km s$^{-1}$ relative to the host star was measured via cross-correlation of a CO molecular template \citep{snellen14}. \cite{haffert19} detected H$\alpha$ around PDS 70 b and c, but the radial velocities measured were of the accretion itself and not of the motion of the planets. \cite{ruffio19} measured the RV of HR 8799 b and c with a 0.5 km s$^{-1}$ precision using a joint forward modeling of the planet signal and the starlight (speckles).

Here we present $R\sim4000$ $K$-band spectra of $\kappa$ And b. In Section \ref{sec:data_red} we report our observations and data reduction methods. In Section \ref{sec:model} we use atmosphere model grids and forward modeling Markov Chain Monte Carlo methods to determine the best-fit effective temperature, surface gravity, and metallicity of the companion.  We use our best-fit parameters and $PHOENIX$ models with scaled molecular mole fractions to derive a C/O ratio of 0.70$_{-0.24}^{+0.09}$ for $\kappa$ And b. In Section \ref{sec:rvs} we use the joint forward modeling technique devised by \cite{ruffio19} to measure $\kappa$ And b's radial velocity and to constrain the plane and eccentricity of its orbit. In Section \ref{sec:discussion} we discuss the implications of our results and future work.  \par

\section{Data Reduction}\label{sec:data_red}
$\kappa$ And b was observed in 2016 and 2017 with the OSIRIS integral field spectrograph (IFS) \citep{Larkin06} in the $K$ broadband mode (1.965--2.381 $\mu$m) with a spatial sampling of 20 milliarcseconds per lenslet.  A log of our observations is given in Table \ref{tab:obslog}. Observations of a blank patches of sky and an A0V telluric standard (HIP 111538) were obtained close in time to the data.  We also obtained dark frames with exposure times matching our dataset. The data were reduced using the OSIRIS data reduction pipeline \citep[DRP;][]{Krabbe04,Lockhart19}. Data cubes are generated using the standard method in the OSIRIS DRP, using rectification matrices provided by the observatory.  At the advice of the DRP working group, we did not use the Clean Cosmic Rays DRP module. We combined the sky exposures from each night and subtracted them from their respective telluric and object data cubes (we did not use scaled sky subtraction). 

\begin{deluxetable}{lccc} 
\tabletypesize{\scriptsize} 
\tablewidth{0pt} 
%\rotate
\tablecaption{OSIRIS Observations of $\kappa$ Andromedae b\label{tab:obslog}} 

\tablehead{ 
  \colhead{Date} & \colhead{Number of} & \colhead{Integration} &
  \colhead{Total Int.} \\
  \colhead{(UT)} & \colhead{Frames} & \colhead{Time (min)} & \colhead{Time (min)}  \\
}
\startdata 
2016 Nov 6 & 5 & 10 & 50  \\
2016 Nov 7 & 8 & 10 & 80  \\
2016 Nov 8 & 5 & 10 & 50  \\
2017 Nov 4 & 13 & 10 & 130  
\enddata
\end{deluxetable}

After extracting one-dimensional spectra for the telluric sources, we used the DRP to remove hydrogen lines, divide by a blackbody spectrum, and combine all spectra for each respective night.  An initial telluric correction for $\kappa$ And b was then obtained by dividing the final combined telluric calibrator spectrum in all object frames.

Once the object data cubes are fully reduced, we identify the location of the planet.  The location can be challenging to find due to the brightness of the speckles even at the separation of the planet ($\sim$1\arcsec\ separation). Speckles have a wavelength-dependent spatial position behavior, and the planet signal does not.  In order to locate the planet, we visually inspect the cubes while stepping through the cube in wavelength, and determine which features do not depend on wavelength.  Once we find the planet, we record the spatial coordinates. 

During preliminary spectral extraction, we noted that the telluric frames did a poor job of correcting some absorption features, particularly in the blue part of the spectrum.  We therefore used the speckles from $\kappa$ And A that are present in all datacubes to derive a telluric correction spectrum for each individual exposure.  This correction works well because $\kappa$ And A is a B9 type star with very few intrinsic spectral lines, so the majority of the spectral features will be from Earths' atmosphere.  We masked the location of the planet and extracted a 1-D spectrum from the rest of the datacube to use as the telluric spectrum.  As with the A0V star, we removed the hydrogen lines and blackbody spectrum based on the temperature of $\kappa$ And A.  

\begin{figure*}
\epsscale{0.60}
\plotone{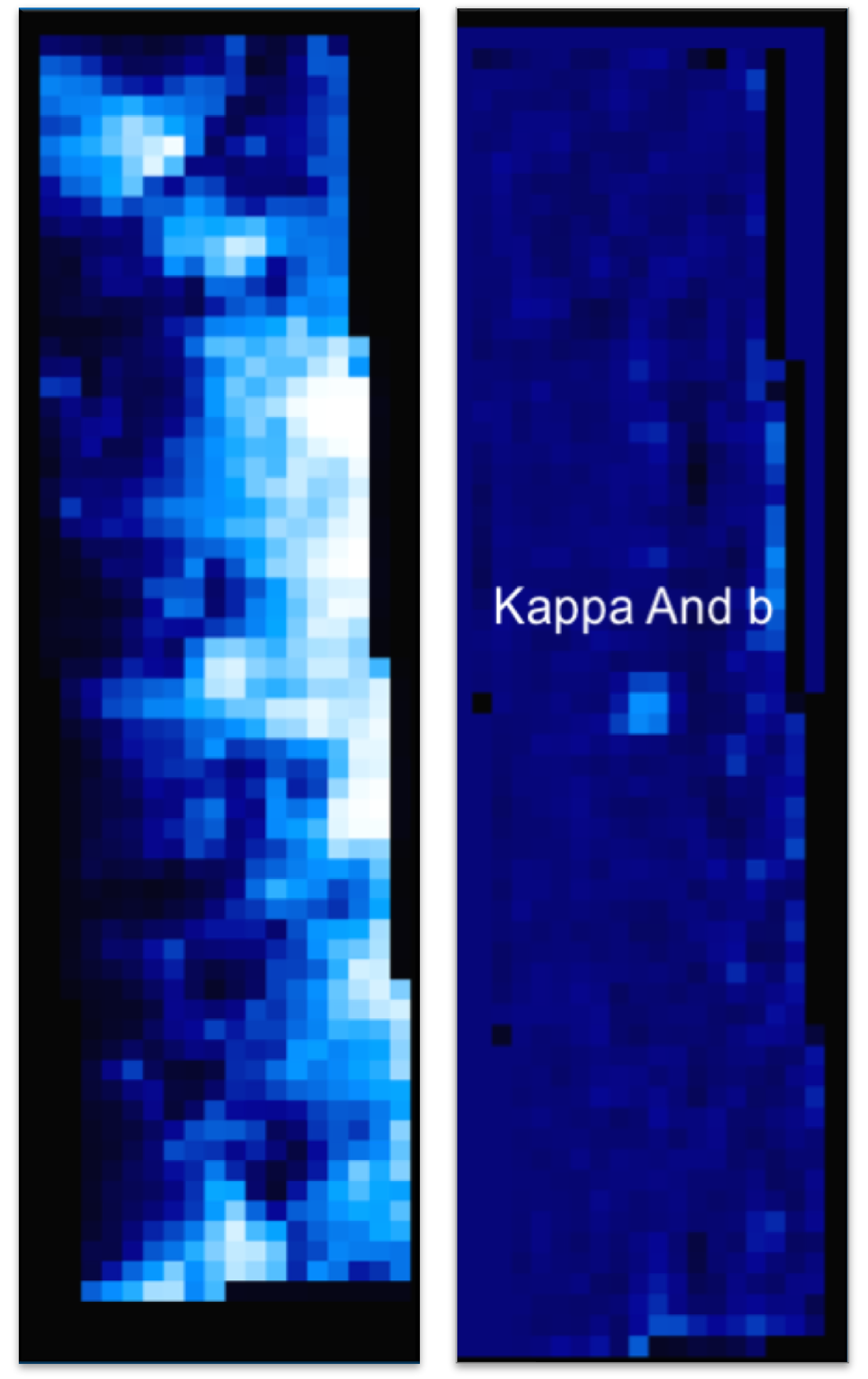}
\caption{An example data cube image frame, collapsed in wavelength via median, from our OSIRIS $\kappa$ And b data set.  The panel on the left shows a reduced cube before speckle removal, demonstrating the brightness of the speckles at the location of $\kappa$ And b.  The right panel shows the data cube after the speckle removal process.  The algorithm effectively removes the speckle noise, leaving most of the flux from the planet behind for spectral extraction.}
\label{fig:speckles}
\end{figure*}

%speckle subtraction
Once the data cubes were reduced and planet location identified, we used a custom IDL routine to remove speckles.  The program smooths and rebins the data to $\lambda/\Delta\lambda\sim$50, and then magnifies each wavelength slice, $\lambda$, about the star by $\lambda m/ \lambda$ with $\lambda m = 2.173~\mu$m, the median wavelength in the $K$-band. The generated data cube has speckles that are positionally aligned, with the planet position varying. The program then fits first order polynomials to every spatial location as a function of wavelength \citep{Barman11,Konopacky13}. We know the position of the planet, and use it to mask the planet to prevent bias in the polynomial fit. The results of the fits are subtracted from the full-resolution spectrum before the slices are demagnified. The resultant cube is portrayed in Figure \ref{fig:speckles} by showing one of the spaxels before and after the speckle removal.

Uncertainties were determined by calculating the RMS between the individual spectra at each wavelength. These uncertainties include contributions from statistical error in the flux of the planet and the speckles as well as some additional error in the blue end of the spectrum due to imperfect removal of large telluric features in this region. The OH sky lines are well-subtracted and have a negligible contribution to the uncertainties. 

We also tested our reduction methodology by planting a fake planet with a flat spectrum in each data cube and going through the same reduction process as above. When we ran the speckle subtraction and then extracted the fake planet spectra from each cube, there were some fluctuations in the spectra, particularly near the ends of the spectral range. We decided to test the speckle subtraction algorithm and extract the fake planet spectra using a higher order polynomoial fit, but the continuum fluctuations were much larger. We therefore determined that the first order polynomial fit introduces the least continuum bias to our data. The uncertainties from the extracted spectra incorporate most of the impact of this bias, with some residual impact at the blue and red ends.  We mitigate the impact in further analysis through removal of the continuum (see Section \ref{sec:temp_g_m}).

\begin{figure*}
\epsscale{0.95}
\plotone{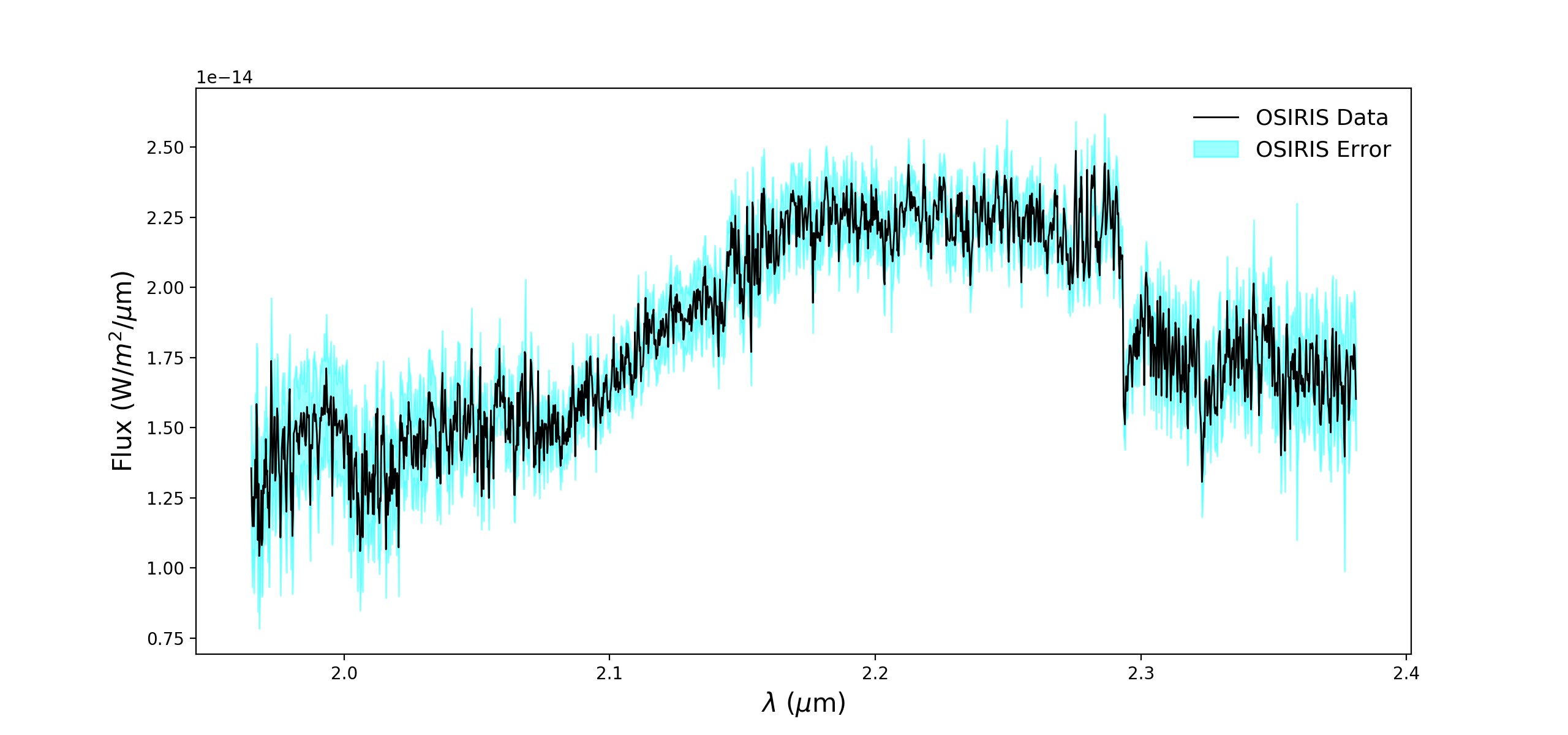}
\caption{Our fully reduced, combined, and flux calibrated moderate-resolution OSIRIS $K$-band spectra of $\kappa$ And b. The errors are shown as a shaded light blue region.}
\label{fig:extracted_spec}
\end{figure*}

Once the speckles are removed, we extract the object spectrum using a box of $3 \times 3$ spatial pixels (spaxels). Once we extracted the $\kappa$ And b spectra from each frame for all data, we then normalize each individual spectrum to account for seeing and background fluctuations, and we apply a barycentric correction to each spectrum. Finally, we median-combine all 30 individual spectra. To calibrate the flux of our spectra we calculated the flux at each wavelength such that, when integrated, the flux matches the $K$-band apparent magnitude ($14.37 \pm 0.07$) from \cite{currie18}.  Figure \ref{fig:extracted_spec} shows the combined, flux calibrated spectrum for $\kappa$ And b.

Once we had our fully reduced, combined, and flux calibrated spectra, we wanted to analyze the spectrum both with and without the continuum. The expectation is that by removing the continuum, some of the residual correlated noise from the speckles get removed as well. To remove the continuum, we apply a high-pass filter with a kernel size of 200 spectral bins to each of the individual spectra. Then we subtract the smoothed spectrum from the original spectra. Once all the individual spectra had been continuum subtracted we median combined them as well, and find the uncertainties by calculating the RMS of the individual spectra at each wavelength.

\section{Spectral Modeling}\label{sec:model}

\subsection{Synthetic Spectra}

Our first goal is to constrain the temperature, surface gravity, and metallicity of $\kappa$ And b.  In order to do this, we must construct a model grid that spans the expected values of these parameters.  In a number of previous works on $\kappa$ And b, the temperature was estimated to be $\sim$2000 K and the surface gravity ($\log g$) $<5$ \citep[e.g.,][]{Hinkley13,Bonnefoy14,todorov16,currie18,uyama20}.  The metallicity has not been constrained, but estimates from the host star suggest values near solar or slightly subsolar \citep{wu11,jones16}.

Based on these measurements, we generated a custom grid based on the \textit{PHOENIX} model framework.   The details on the computation of this grid are described in \cite{Barman11,barman15}, with the updated methane line list from \cite{yurchenko14} and the optical opacities from \cite{karkoschka10}.  The grid spans a temperature range 1500--2500~K, a $\log g$ range of 2--5.5~dex, and a metallicity range of $-0.5$--0.5~dex, which encompasses the range of values previously reported for $\kappa$ And b. For a $\sim$2000~K object, the C is already in CO instead of CH$_4$ throughout the atmosphere, and thus the amount of CO should be constant with height. Therefore, for $\kappa$ And b, we chose not to model vertical mixing (Kzz = 0).  

The cloud properties for young gas giants and brown dwarfs are notoriously complex.  In our modeling framework, we are able to incorporate clouds in several different ways.  We can generate a thick cloud with an ISM-like grain size distribution (DUSTY, \citealt{allard01}), a complete lack of cloud opacity (COND, \citealt{allard01}), or an intermediate model that spans these two extremes (ICM, \citealt{Barman11}).  Given the estimated temperature and surface gravity of $\kappa$ And b, we chose to use a $(DUSTY)$ cloud model in our grid, which has been shown to do a reasonably good job at reproducing brown dwarf spectra with similar properties \citep[e.g.,][]{kirkpatrick06}.  We will therefore refer to the custom grid constructed here as \textit{PHOENIX-ACES-DUSTY} to distinguish it from other models based on the \textit{PHOENIX} framework.  We explore the results of this choice of cloud model and describe the results from a few other models in Section \ref{sec:temp_g_m}.       

The synthetic spectra from the grid were calculated with a wavelength sampling of 0.05~\AA\ from 1.4 to 2.4~$\mu$m. Each spectrum was convolved with a Gaussian kernel with a FWHM that matched the OSIRIS spectral resolution \citep{barman15}. Both flux calibrated and continuum subtracted data were modeled and analyzed. The synthetic spectra was flux calibrated and continuum subtracted using the same routines as the data.

\subsection{Forward Modeling}\label{sec:for_mod}

To determine the best-fit \textit{PHOENIX-ACES-DUSTY} model, we use a forward-modeling approach following \cite{Blake2010}, \cite{Burgasser16}, Hsu et al. (in prep), and Theissen et al. (in prep). The effective temperature ($T_\mathrm{eff}$), surface gravity ($\log g$), and metallicity ([M/H]) are inferred using a Markov Chain Monte Carlo (MCMC) method built on the \texttt{emcee} package that uses an implementation of the affine-invariant ensemble sampler \citep{GoodmanWeare10,ForemanMackey13}. 

We assume that each parameter we are solving for should be normally distributed, and thus the log-likelihood function is computed as follows
\begin{equation}
\ln L = -0.5 \times \left[\sum \left[{\frac{\mathrm{data}[p] - D[p]}{\sigma[p]}}\right]^2 + \\ \sum ln(2\pi{\sigma}^2)\right],
\end{equation}
where \(\sigma\) is the provided uncertainties, data\([p]\) is our science data, and \(D[p]\) is the forward-modeled data. The uncertainty is taken as the difference between the 84th and 50th percentile as the upper limit, and the difference between the 50th and 16th percentile as the lower limit for all model parameters. If the posterior distributions follow normal (Gaussian) distributions then this equates to the 1-\(\sigma\) uncertainty in each parameter (e.g., \citealt{Blake2010, Burgasser16}).  Assuming that there are no additional systematic uncertainties in the data or in the models, these uncertainties should be an accurate reflection of our knowledge of each parameter.  We discuss and attempt to account for additional systematic uncertainties in the data and the models in Section \ref{sec:temp_g_m}.

The data is forward-modeled using the following equation:
\begin{equation}
D[p] = C \times \left[\left(M\left[p\left(\lambda \left[1 - \frac{RV}{c}\right]\right),T_\mathrm{eff},\log g, [M/H]\right]\right)\right] * \kappa_G(\Delta \nu_\mathrm{inst}) + C_\mathrm{flux}.
\end{equation}
Here, \(p(\lambda)\) is the mapping of the wavelength values to pixels, \(M[p(\lambda)]\) is the stellar atmosphere model parameterized by effective temperature ($T_\mathrm{eff}$), surface gravity ($\log g$), and metallicity ([M/H]), \(C\) is the dilution factor, (radius/distance)$^2$, that scales the model to the observed fluxes, (which is measure of radius since the distance is known, e.g., \citealt{theissen14,kesseli19}), and \(\kappa_G(\Delta \nu_\mathrm{inst})\) is the line spread function (LSF) calculated from the OSIRIS resolution of $R = 4000$ to be 34.5~km s$^{-1}$. The \(RV\) is the radial velocity that is used here only to account for wavelength calibration errors in the OSIRIS DRP, \(c\) is the speed of light, and \(C_\mathrm{flux}\) is an additive continuum correction to account for potential systematic offsets in the continuum. This final parameter (\(C_\mathrm{flux}\)) is only used when fitting the continuum normalized data.  Our MCMC runs used 100 walkers, 500 steps, and a burn-in of 400 steps to ensure parameters were well mixed.

\subsection{Temperature, Gravity, and Metallicity}\label{sec:temp_g_m}

\begin{figure*}
\epsscale{0.95}
\plotone{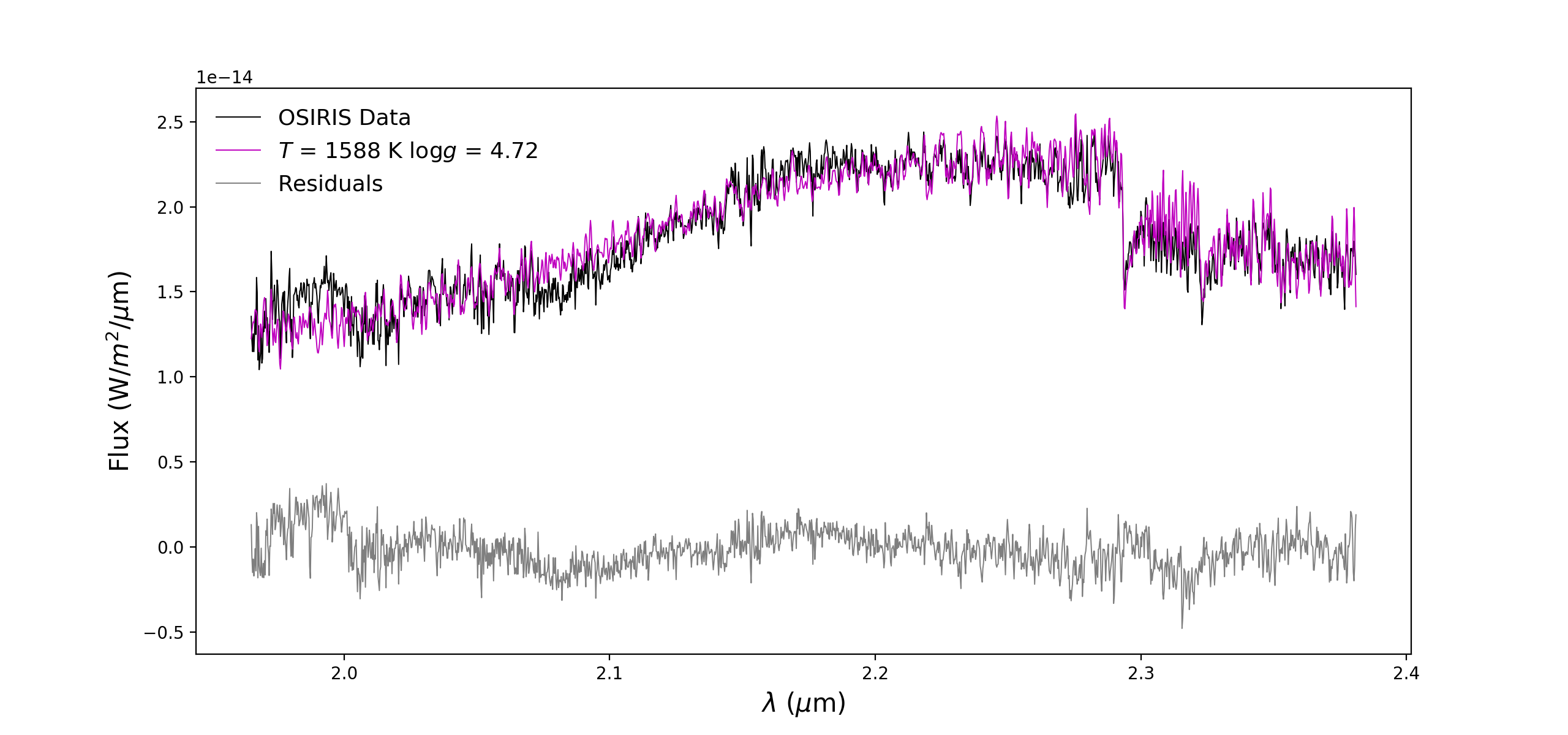}
\caption{Results from the MCMC model fit to the OSIRIS spectrum for $\kappa$ And b without continuum removal (black).  The best matching \textit{PHOENIX-ACES-DUSTY} model has $T_\mathrm{eff}$ = 1588 K, $\log g$ = 4.72, and a [M/H] = $0.50$ (magenta). The residuals between the data and the model are plotted in gray.  The shape of the continuum is impacted by speckle noise, which modulates at low spatial frequencies and leaves residual noise in our dataset post speckle removal.  The fits are then driven to lower temperatures and higher metallicities than previously found for $\kappa$ And b due to the continuum shape impact, particular at blue wavelengths.}
\label{fig:model_continuum}
\end{figure*}

\begin{figure*}
\epsscale{0.95}
\plotone{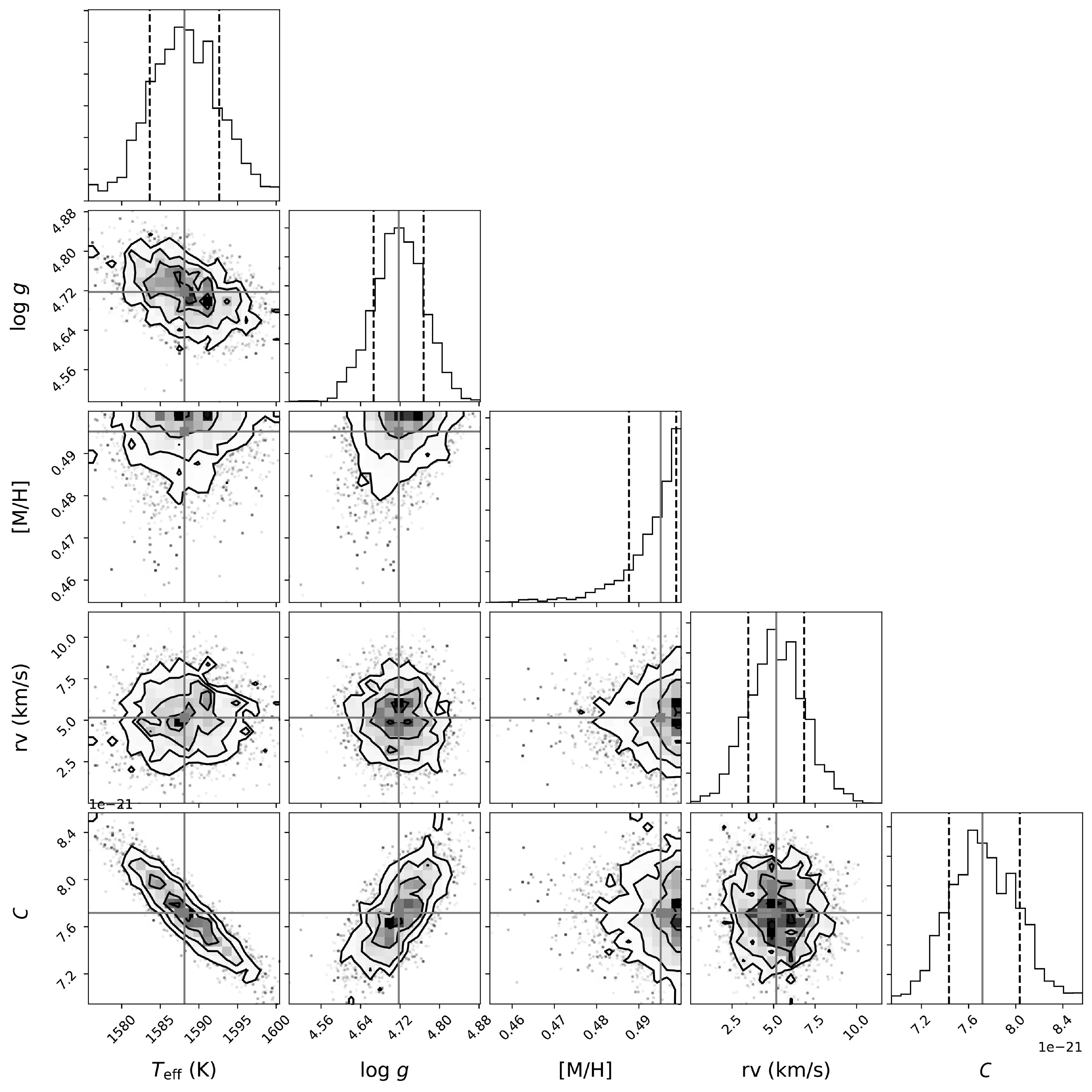}
\caption{Corner plot corresponding to the fit shown in Figure \ref{fig:model_continuum}. \{The diagonal shows the marginalized posteriors. The subsequent covariances between all the parameters are in the corresponding 2-d histograms. The blue lines represent the 50 percentile, and the dotted lines represent the 16 and 84 percentiles. $C$ corresponds to the dilution factor that scales the model by \((radius)^2 (distance)^{-2}\) as mentioned in Section \ref{sec:for_mod}}.
\label{fig:corner_continuum}
\end{figure*}

\begin{figure*}
\epsscale{0.95}
\plotone{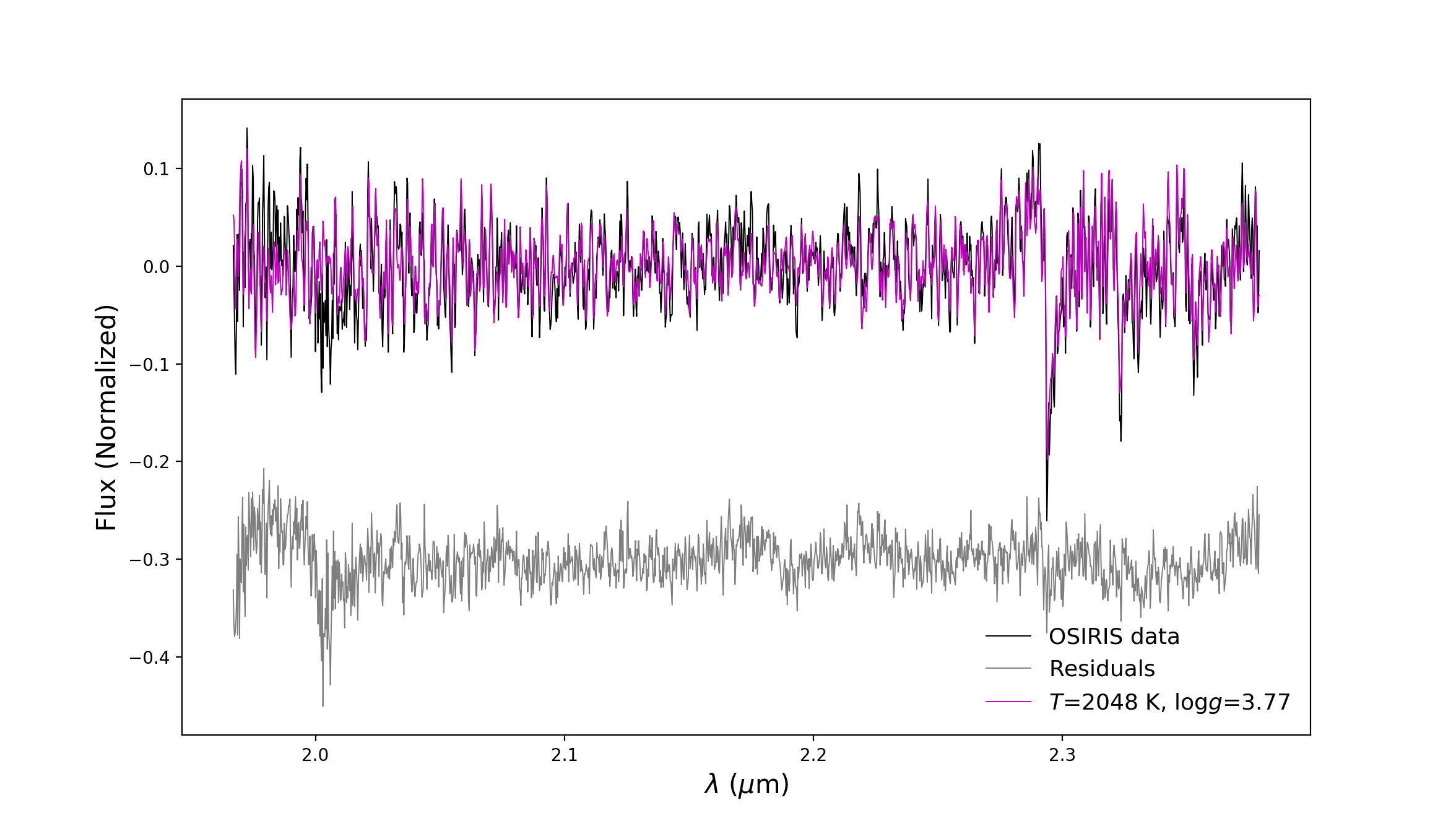}
\caption{Results from MCMC model fit to the OSIRIS spectrum after continuum removal (black).  The best fitting model has  $T_\mathrm{eff}$ = 2048 K, $\log g$ = 3.77, and [M/H] = $-0.11$ in magenta. The residuals between the flattened data and the flattened model are in gray.}
\label{fig:model_flat}
\end{figure*}

\begin{figure*}
\epsscale{0.95}
\plotone{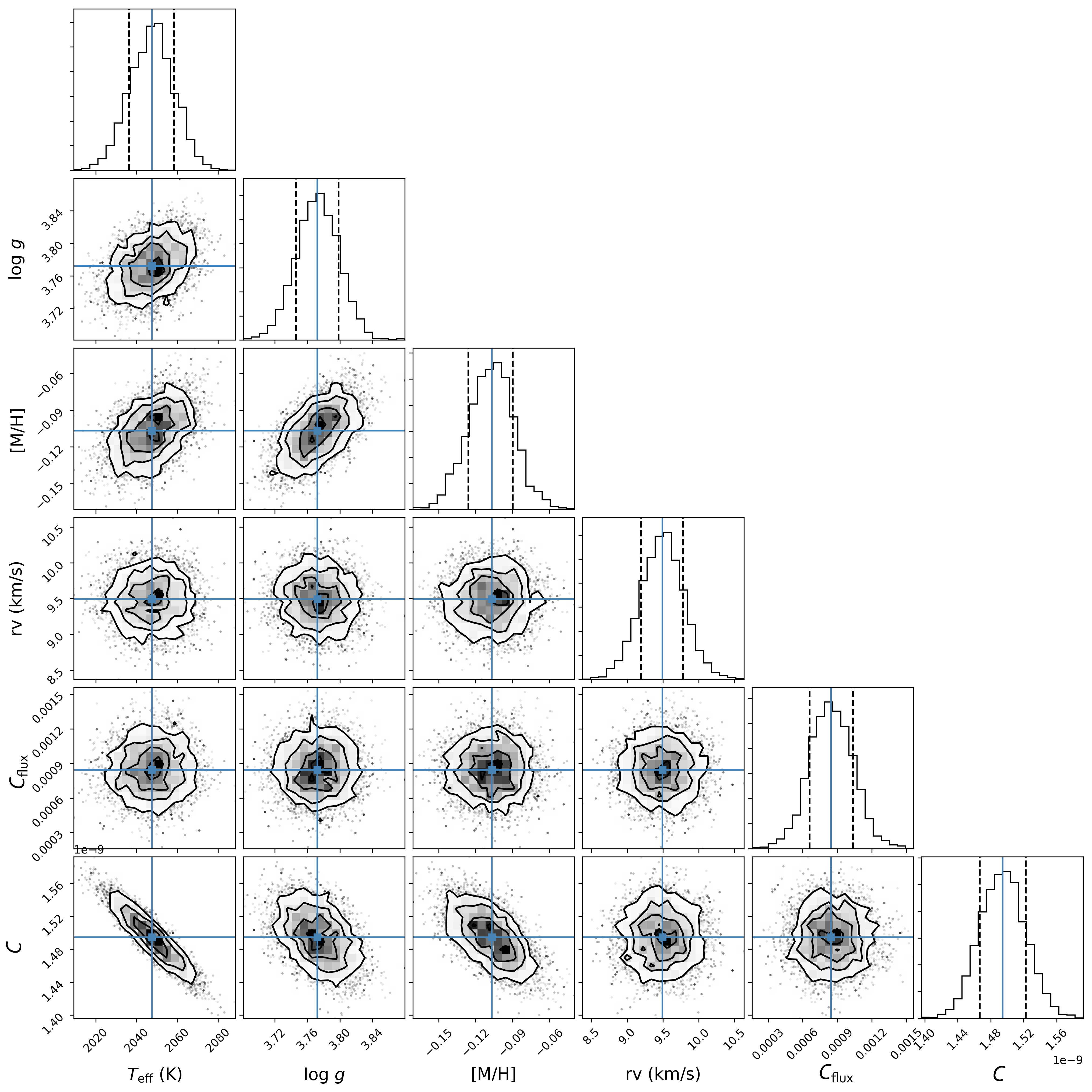}
\caption{Corner plot corresponding to the fit shown in Figure \ref{fig:model_flat}. The diagonal shows the marginalized posteriors. The subsequent covariances between all the parameters are in the corresponding 2-d histograms. The blue lines represent the 50 percentile, and the dotted lines represent the 16 and 84 percentiles. $C_\mathrm{flux}$ is the additive flux parameter and $C$ corresponds to the dilution factor that scales the model by \((radius)^2 (distance)^{-2}\) as mentioned in Section \ref{sec:for_mod}}.
\label{fig:corner_flat}
\end{figure*}

We ran our MCMC fitting procedure on both the flux calibrated spectrum and the continuum-subtracted spectrum. The best-fit parameters for our flux calibrated data are $T_\mathrm{eff} = 1588 \pm 5$ K, $\log g = 4.72^{+0.05}_{-0.06}$, and a metallicity of [M/H] = $0.5 \pm 0.01$.  For the radius, which comes from the multiplicative flux parameter, we found R = 1.00 $\pm$ 0.02 R$_{Jup}$.  For our continuum-subtracted data the best-fit parameters were $T_\mathrm{eff} = 2048 \pm 11$~K, $\log g = 3.77 \pm 0.03$, and a metallicity of [M/H] = $-0.11 \pm 0.02$. Radii cannot be derived for the continuum-subtracted data. Figures \ref{fig:model_continuum} through \ref{fig:corner_flat} show the best-fit spectrum overplotted on our data, and the resulting corner plots from our MCMC analysis for both the initially extracted and continuum-subtracted spectra.

\begin{deluxetable*}{lccccc} 
\tabletypesize{\scriptsize} 
\tablewidth{0pt} 
%\rotate
\tablecaption{Summary of atmospheric parameters derived from MCMC fits.}
\label{tab:atm_param}
\tablehead{ 
  \colhead{Spectra} & \colhead{Effective Temperature} & \colhead{Surface Gravity} &
  \colhead{Metallicity} & \colhead{Radius} & \colhead{Luminosity}\\
  \colhead{$\kappa$ And b} & \colhead{${T}_\mathrm{{eff}}$ (K)} & \colhead{$\log g$} & \colhead{[M/H]} &\colhead{($R_\mathrm{Jup}$)} &\colhead{$\log_{10}\left(\frac{L}{L_\odot}\right)$} 
}
\startdata 
\multicolumn{6}{c}{PHOENIX-ACES-DUSTY} \\
\hline
OSIRIS Including Continuum & $1588 \pm 5$ & $4.72^{+0.05}_{-0.06}$ & $0.50^{+0.01}_{-0.01}$ & $1.0 \pm 0.02$ & $-4.2 \pm 0.1$ \\
OSIRIS Continuum Subtracted & $2048 \pm 11$  & $3.77 \pm 0.03$ & $-0.11 \pm 0.02$ & n/a & n/a\\
CHARIS All Bands  & $2021^{+20}_{-19}$ & $3.64^{+0.18}_{-0.10}$ & $0.46^{+0.03}_{-0.07}$ & $0.99 \pm 0.02$ & $-3.80 \pm 0.02$\\
CHARIS K Band Only & $1707^{+147}_{-118}$ & $4.62^{+0.48}_{-0.63}$ & $-0.12^{+0.28}_{-0.23}$ & $1.4 \pm 0.2$ & $-3.8 \pm 0.1$ \\
SpeX 2MASS J01415823$-$4633574 & $1972^{+9}_{-10}$ & $2.93^{+0.08}_{-0.14}$ & $0.49 \pm 0.01$ & n/a & n/a \\
\hline
\multicolumn{6}{c}{BT-SETTL} \\
\hline
OSIRIS Including Continuum & $1630^{+7}_{-5}$ & $3.5^{+0.5}_{-0.4}$ & n/a & $1.10 \pm 0.10$ & $-4.11 \pm 0.1$ \\
OSIRIS Continuum Subtracted & $2128^{+70}_{-73}$ & $4.47^{+0.02}_{-0.06}$ & n/a & n/a & n/a \\
CHARIS All Bands & $1817^{+48}_{-18}$ & $5.15^{+0.13}_{-1.13}$ & n/a & $1.2 \pm 0.1$ & $-3.8 \pm 0.1$\\
CHARIS K Band Only & $1647^{+194}_{-96}$ & $4.16^{+0.35}_{-0.33}$ & n/a & $1.5 \pm 0.3$ & $-3.8 \pm 0.2$\\
\hline
\multicolumn{6}{c}{DRIFT-PHOENIX} \\
\hline
OSIRIS Including Continuum & $2200^{+100}_{-130}$ & $4.0^{+0.3}_{-0.5}$ & n/a & $1.0^{+0.2}_{-0.1}$ & $-3.6 \pm 0.2$ \\
OSIRIS Continuum Subtracted & $2126^{+104}_{-131}$ & $4.19^{+0.2}_{-0.22}$ & n/a & n/a & n/a \\
CHARIS All Bands & $1747^{+20}_{-18}$ & $3.99^{+0.19}_{-0.20}$ & n/a & $1.5 \pm 0.1$ & $-3.7 \pm 0.1$\\
CHARIS K Band Only & $1863^{+289}_{-233}$ & $4.22^{+0.50}_{-0.46}$ & n/a & $1.3 \pm 0.3$ & $-3.7 \pm 0.2$ \\
\hline
Adopted Values & 2050 & 3.8 & -0.1 & 1.2 & -3.8 \\
Range of Allowed Values & 1950 - 2150 & 3.5 - 4.5 & -0.2 - 0.0 & 1.0 - 1.5 & -3.5 - -3.9
\enddata
\tablenotetext{-}{The grid used in each case is noted above derived parameters.  Using the range of best-fit values and our estimates of systematic uncertainties, range of adopted atmospheric parameters for $\kappa$ And b are shown in the last row. We are using the convention for metallicity where [M/H] = $\log _{10}\left(\frac{N_{M}}{N_{H}}\right)_\mathrm{star} - \log _{10}\left(\frac{N_{M}}{N_{H}}\right)_{\sun}$}
\end{deluxetable*}

The discrepancy between the two fits, one with the continuum and one without, is not entirely unexpected.  The continuum is strongly impacted by residual systematic errors from the speckle noise, which injects features at low spatial frequencies.  Effective temperature is particularly sensitive to continuum shape, and as a bolometric quantity is better estimated by including data from a broader range of wavelengths.  Subtracting the continuum mitigates and removes some of these residual errors.

%\FloatBarrier

\begin{figure*}
\epsscale{0.95}
\plotone{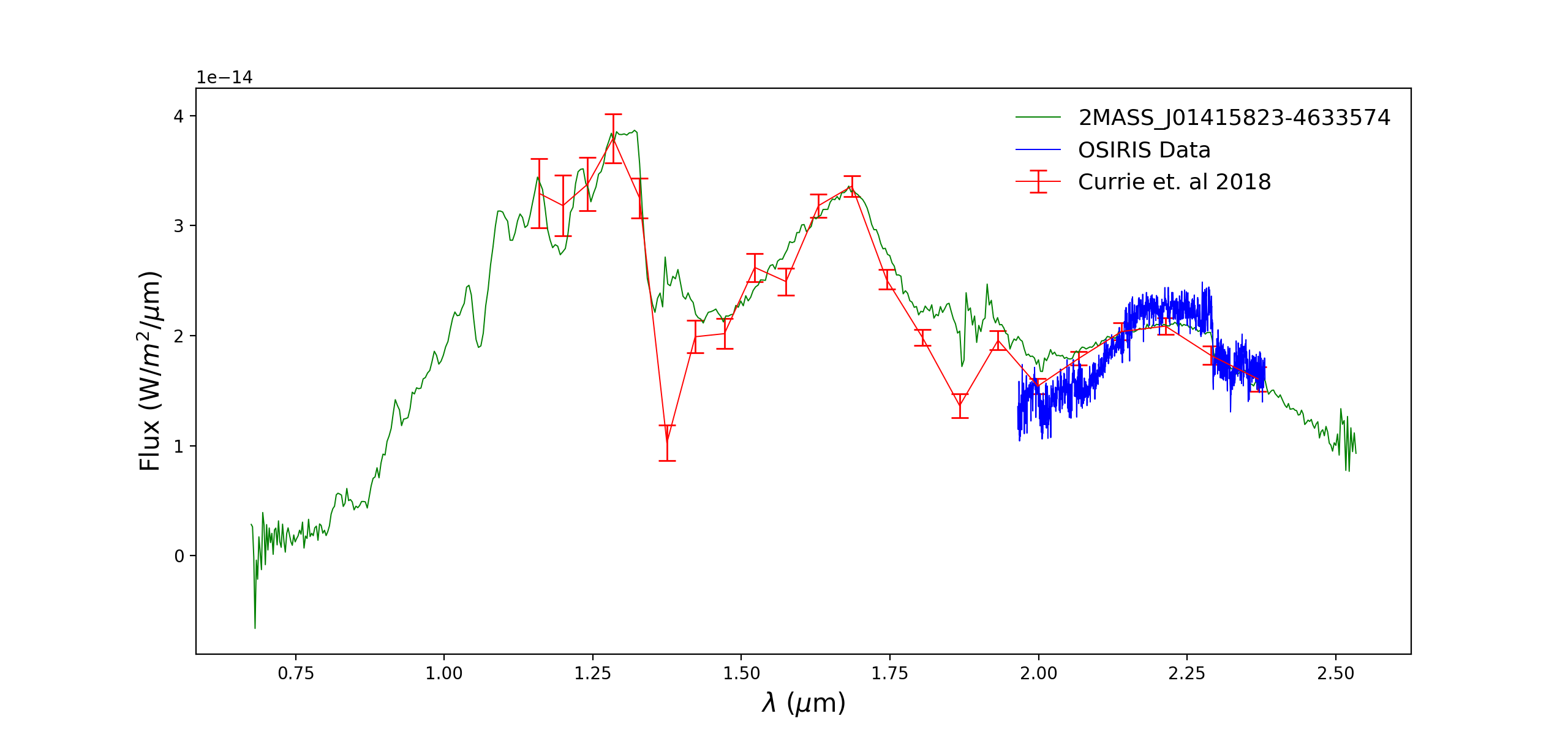}
\caption{OSIRIS $K$-band data of $\kappa$ And b compared to \cite{currie18} low-resolution CHARIS data of $\kappa$ And b and their best-matching field source, 2MASS J01415823-4633574 from the SpeX Library \citep{kirkpatrick06}.  A fit to the SpeX spectrum (not shown) reveals temperatures and gravities consistent with the OSIRIS and CHARIS data on $\kappa$ And b.}
\label{fig:charis_2mass}
\end{figure*}

In order to verify that the temperature estimates we derived from the flattened spectra are robust, we ran our MCMC fitting code using the \textit{PHOENIX-ACES-DUSTY} grid on the CHARIS spectrum from \cite{currie18}, which spans a much larger range of wavelengths (Figure \ref{fig:charis_2mass}).  We adjusted our MCMC code for the CHARIS data by changing the LSF to 7377~km s$^{-1}$ for the instrument.  We fit all near-infrared bands simultaneously, and also performed a fit using only the $K$-band. For the fit to all the bands simultaneously, we obtained $T_\mathrm{eff} = 2021^{+20}_{-19}$~K, $\log g = 3.64^{+0.18}_{-0.10}$, [M/H] = $0.46^{+0.03}_{-0.07}$, and R = 0.99 $\pm$ 0.02.  When we fit only the $K$-band of the CHARIS spectrum we obtained $T_\mathrm{eff} = 1707^{+147}_{-118}$~K, $\log g = 4.62^{+0.48}_{-0.63}$,  [M/H] = $-0.12^{+0.28}_{-0.23}$, and R=1.4$\pm$0.2. The all-wavelength fit is consistent with the results we obtained for our continuum-normalized spectrum fitting of the OSIRIS data, while K-band only is slightly lower in temperature, albeit with large uncertainties.  Our fits to the CHARIS data are also consistent with the results obtained in \citet{currie18} and \citet{uyama20} ($T_\mathrm{eff} \approx 1700--2000~K, \log g \approx 4--4.5$, R=1.3--1.6 R$_{Jup}$).  For a more detailed comparison of the OSIRIS continuum to the CHARIS spectrum, we binned our OSIRIS $K$-band spectra to the same sampling as the \cite{currie18} spectra shown in Figure~\ref{fig:charis_2mass}. The spectra were consistent except for the OSIRIS spectral peak was shifted very slightly towards the red. Two CHARIS data points are less than 1.5-$\sigma$ off from our OSIRIS data, and the rest (4 additional points) are consistent within the error bars. 

Figure~\ref{fig:charis_2mass} also shows a comparison between our spectrum and the best-matching brown dwarf from the SpeX prism library \citep{Burgasser14} found by \citet{currie18}, 2MASS J01415823-4633574 \citep{kirkpatrick06}.  This source is a young, early L-type object associated with the Tucana-Horologium association (age $\sim 40$~Myr).  Since the match to this brown dwarf is quite good, we also fit its spectrum using the same model grid and our MCMC framework, adjusting the model resolution to match SpeX.  We found fully consistent properties with temperatures between $\sim$2050--2130~K and a $\log g \approx 3$--4 for this source.

\begin{figure*}
\epsscale{0.95}
\plotone{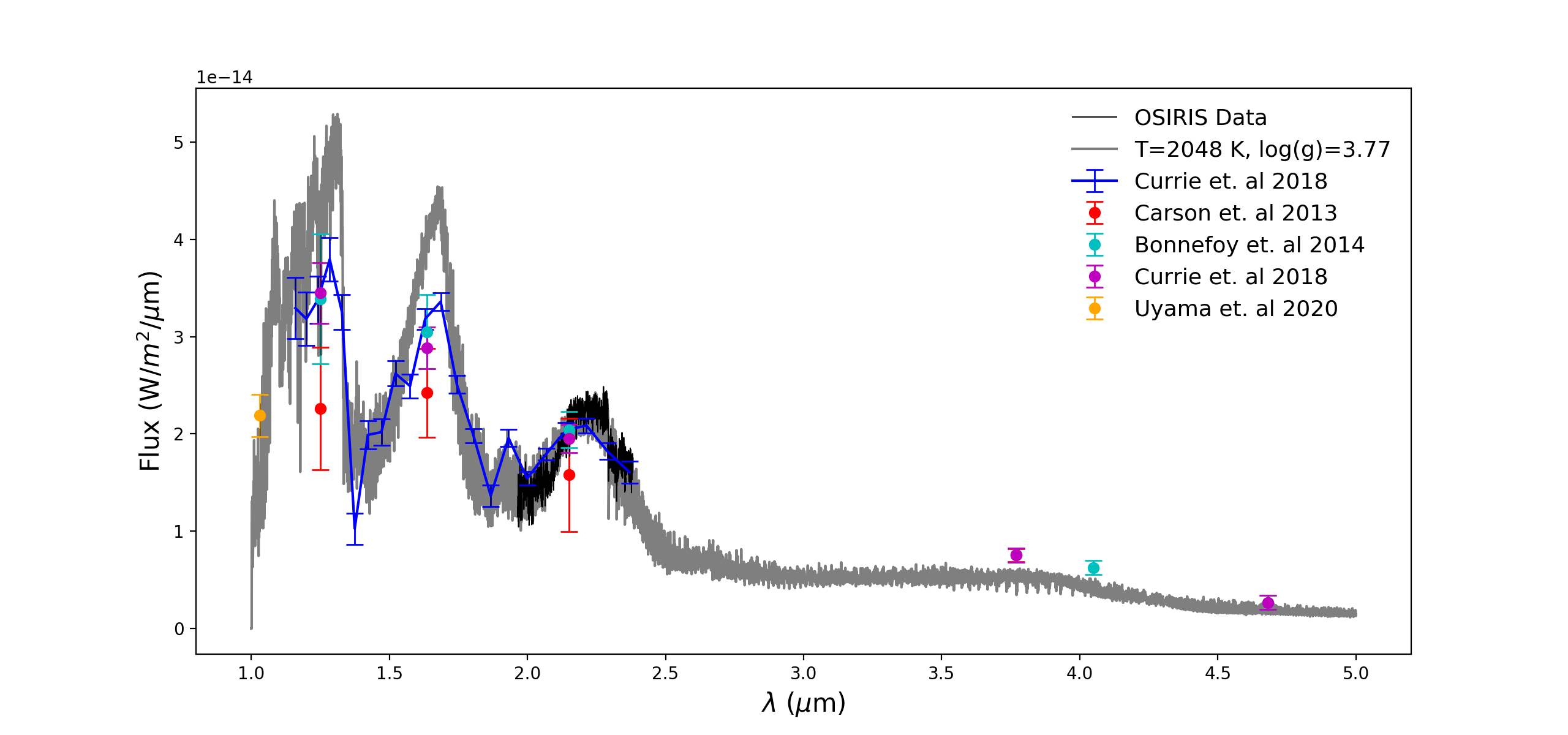}
\caption{All available spectral and photometric data for $\kappa$ And b compared to the best-fit \textit{PHOENIX-ACES-DUSTY} model, shown in gray, of $T_\mathrm{eff}$ = 2048 K, $\log g$ = 3.77, and M/H = $-0.11$ over the near-infrared. Our OSIRIS data are shown in black. \cite{currie18} low-resolution CHARIS spectra is plotted in dark blue. Photometric data points are taken from \cite{Bonnefoy14}, \cite{Carson13}, \cite{currie18}, and \cite{uyama20}.  The model matches the data at $K$-band, but predicts higher flux in $H$- and $J$-band (though the morphology is consistent).  The mismatch at the low and high wavelength range is likely due to our use of a DUSTY cloud model.}
\label{fig:bigplot}
\end{figure*}

Figure~\ref{fig:bigplot} shows all available spectral and photometric data for $\kappa$ And b.  Overplotted on the spectrum is the best-fit \textit{PHOENIX-ACES-DUSTY} model based on the continuum-subtracted OSIRIS spectrum.  The model is scaled to match the continuum flux at $K$-band, which in turn is derived using the $K$-band magnitude in \citet[$K_s = 14.37 \pm 0.07$;][]{currie18}.  The match to the \citet{currie18} spectrum is quite good, in alignment with the consistent effective temperature we derive from fitting that data set with our models.  While the shape of the $J$- and $H$-band spectra is similar to the CHARIS spectrum, the model over-predicts the flux by 2--7-$\sigma$ in the $H$-band and 1--4-$\sigma$ in the $J$-band, and slightly underpredicts the flux by $\sim$1.5-$\sigma$ near 4 $\mu$m.  The reason that a similar temperature is derived from an all-band fit to the CHARIS spectrum using our models is that the flux scaling parameter, and thus the radius, is lowered in this case such that it results in the model ``trisecting" the three wavelengths, matching $J$- and $H$-band quite nicely, but then underpredicting the $K$-band flux.

The mismatch at $J$ and $H$ bands could almost certainly be due to the cloud properties used in our grid.  We are using a DUSTY cloud model, which is meant to be a limiting case of a true thick cloud model.  Generally, DUSTY models do a reasonable job at matching spectra in this temperature range (2000--2500~K).  A slight modification to the cloud properties could result in a general change to the flux at a given band without dramatically impacting the spectral morphology.  Given the insensitivity of the continuum-normalized OSIRIS spectrum to clouds, it is encouraging that all fits are returning consistent temperatures in spite of the flux offsets.

A recent analysis of the CHARIS data by \citet{uyama20} found a slightly lower temperature using models from \citet{allard12}, \citet{Chabrier00}, and \citet{witte11} (\textit{BT-SETTL}, \textit{BT-DUSTY}, \textit{DRIFT-PHOENIX}).  These models have different assumptions about cloud properties than we used in our grid. The \textit{BT-SETTL} grids treat clouds with number density and size distribution as a function of depth based on nucleation, gravitational settling, and vertical mixing \citep{allard12}. The \textit{DRIFT-PHOENIX} grids treat clouds by including effects of nucleation, surface growth, surface evaporation, gravitational settling, convection, and element conservation \citep{witte11}. \citet{uyama20} were able to get very good matches at all wavelengths using these models, with temperatures of 1700--1900~K and $\log g$ between 4--5.  The range of uncertainties they found encompasses $\sim$2000~K, and were close to the range of temperatures we find with \textit{PHOENIX-ACES-DUSTY}.  

Since our subsequent analysis of the chemical abundances of $\kappa$ And b relies on knowledge of the temperature and gravity, we did additional modeling to look at the comparison between these models and our continuum-normalized OSIRIS data.  In addition to differences in cloud parameters, each set of models incorporates slightly different assumptions that lead to systematic differences in the output spectra for the same parameters such as temperature and gravity (e.g., \citealt{oreshenko20}).  These systematics are not captured in the formal uncertainties from each MCMC run.  We attempt to account for these systematics by looking at the range of values given from the three models.  

We incorporated both the \textit{BT-SETTL} and \textit{DRIFT-PHOENIX} models into our MCMC analysis code, and fit our OSIRIS spectrum using the same procedure described above.  The best-fit using \textit{BT-SETTL} yielded $T_\mathrm{eff} = 2128^{+70}_{-73}$ K and $\log g = 4.47^{+0.02}_{-0.06}$.  The \textit{DRIFT-PHOENIX} models generally provided poor matches to the higher resolution data, but yielded $T_\mathrm{eff} = 2126^{+104}_{-131}$ K and $\log g = 4.19^{+0.2}_{-0.22}$ as best-fit parameters.  We found no fits with \textit{DRIFT-PHOENIX} that properly captured the first drop of the CO bandhead at $\sim$2.9 $\mu$m.  We also fit the CHARIS data using our code and these model grids, and found parameters consistent with \citet{uyama20}. We then looked in detail at the difference between our best-fits to the OSIRIS data and these lower temperature models at $R \sim 4000$.  The $\chi^2$ of the best fits ($T_\mathrm{eff} = 2100$~K) is significantly better than the $\chi^2$ of a $T_\mathrm{eff} =1700$~K, $\log g = 4$ model, by roughly 5$\sigma$ using either grid.

Table 2 shows the results for all atmospheric parameters derived in this work.  We use the range of best-fit values from the OSIRIS continuum-normalized data to define the adopted parameters for temperature, gravity, and metallicity, as the resolved line information offers the most constraints on those parameters.  We adopt values a value of $T_\mathrm{eff} = 2050$~K, with a range of 1950--2150~K, $\log g=3.8$, with a range of 3.5--4.5, and  [M/H] = $-0.1$, with a range of -0.2--0.0.  For radius, we use the median value from the OSIRIS continuum-included data and the CHARIS data to arrive at $R=1.2~R_{Jup}$, with a range of 1.0--1.5~$R_{Jup}$.  This yields an implied bolometric luminosity of log(L/L$_{\odot}$) = $-3.7$, with a range of $-3.5$ to $-3.9$, consistent with the estimate from \citet{currie18}.  While it is possible that lower temperatures could be invoked for $\kappa$ And b, a more detailed analysis including a variation of cloud models will be required to determine whether this is a viable solution that also matches the OSIRIS data.  Since our high resolution data is not particularly informative for cloud properties, we leave such analysis to future work.  

\subsection{Mole Fractions of CO and H$_2$O}

With best-fit values for temperature, surface gravity, and metallicity we can fit for abundances of CO and H$_2$O in our OSIRIS $K$-band spectra. Once best-fit values were determined for $T_\mathrm{eff}$, $\log g$, and [M/H], we fixed those parameters to generate a grid of spectra with scaled mole fractions of the molecules for the $K$-band \citep{barman15}.  Since our best-fit metallicity was slightly subsolar (roughly 80\% of the solar value), we note that the overall abundances of these molecules will be slightly less than that of the Sun, but their unscaled \textit{ratios} will match the Sun.  The molecular abundances of CO, CH$_4$, and H$_2$O were scaled relative to their initial values from 0 to 1000 using a uniform logarithmic sampling, resulting in 25 synthetic spectra. We fit for the mole fraction of H$_2$O first, holding CO and CH$_4$ at their initial values. The fit was restricted to wavelengths less than the CO band head to avoid biasing from overlapping CO. Next, the H$_2$O mole fraction was set to its nominal value, and we fit for scaled CO.  While in principle we could do the same analysis for CH$_4$, we did not do so because in this temperature regime there is no expectation of a significant amount of CH$_4$ present in our $K$-band spectrum.

\begin{figure*}
\epsscale{0.95}
\plotone{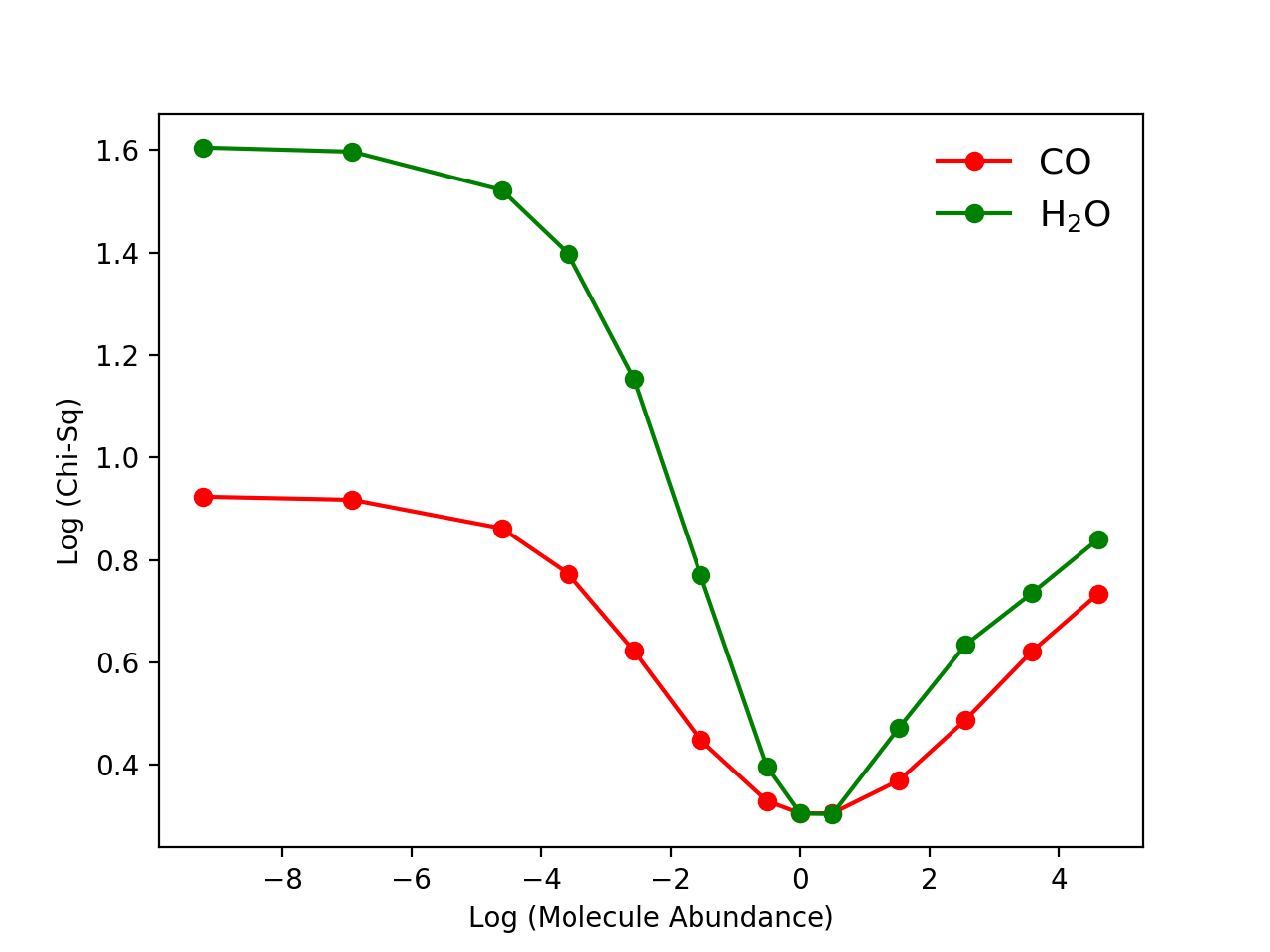}
\caption{Results of $T_\mathrm{eff}$ = 2048 K and $\log g$ = 3.77 model fits with varying mole fractions for both H$_2$O and CO to our continuum-subtracted OSIRIS spectrum.  The mole fractions are given in units relative to the ratio in the Sun, such that a value of zero implies the solar value.  Both scalings of CO and H$_2$O prefer values near solar.  From these fits we find C/O = 0.70$_{-0.24}^{+0.09}$.} 
\label{fig:chisq_co_h2o}
\end{figure*}

Figure \ref{fig:chisq_co_h2o} shows the resulting \(\chi^2\) distribution as a function of CO and H$_2$O mole fraction.  
The models with the lowest \(\chi^2\) when compared to the flattened data gave us the best-fits for both H$_2$O and CO. The best fit for H$_2$O had a scaling of 1, and the best fit for CO had a scaling of 1.66.  To calculate the 1-$\sigma$ uncertainties in each mole fraction value, we used the values from models within $\pm$1 of our lowest \(\chi^2\). Using interpolation along the curves shown in Figure \ref{fig:chisq_co_h2o}, the range of mole fractions encompassed by these uncertainties is 0.599 to 3.24 times the initial mole fraction of CO, and 0.599 to 1.791 times the initial H$_2$O mole fraction. 

\cite{todorov16} derived a water abundance for $\kappa$ And b using spectral retrieval with a one-dimensional plane-parallel atmosphere and a single cloud layer that covers the whole planet. This modeling was done on the low-resolution spectrum from P1640 presented in \cite{Hinkley13}. They derived the $\log(n_\mathrm{H_{2}O})$ for four cases that varied in the treatment of molecular species and clouds.  In each case, they found consistent values for the mole fraction of water, with $\log(n_\mathrm{H_{2}O})$ $\sim-$3.5.  Our best-matching mole fraction for water is $\log(n_\mathrm{H_{2}O})$ $\sim-$3.7, which is consistent within the uncertainties in \cite{todorov16}.  

\subsection{C/O Ratios}
For giant planets formed by gravitational instabilities, their atmospheres should have element abundances that match their host stars \citep{helled2009}. If giant planets form by a multi-step core accretion process, it has been suggested that there could be a range of elemental abundances possible \citep{oberg2011,madhu19}. In this scenario, the abundances of giant planets' atmospheres formed by core/pebble accretion are highly dependent on the location of formation relative to CO, CO$_2$, and H$_2$O frost lines and the amount of solids acquired by the planet during runaway accretion phase.  This can be diagnosed using the C/O ratio. 

\begin{figure*}
\epsscale{0.95}
\plotone{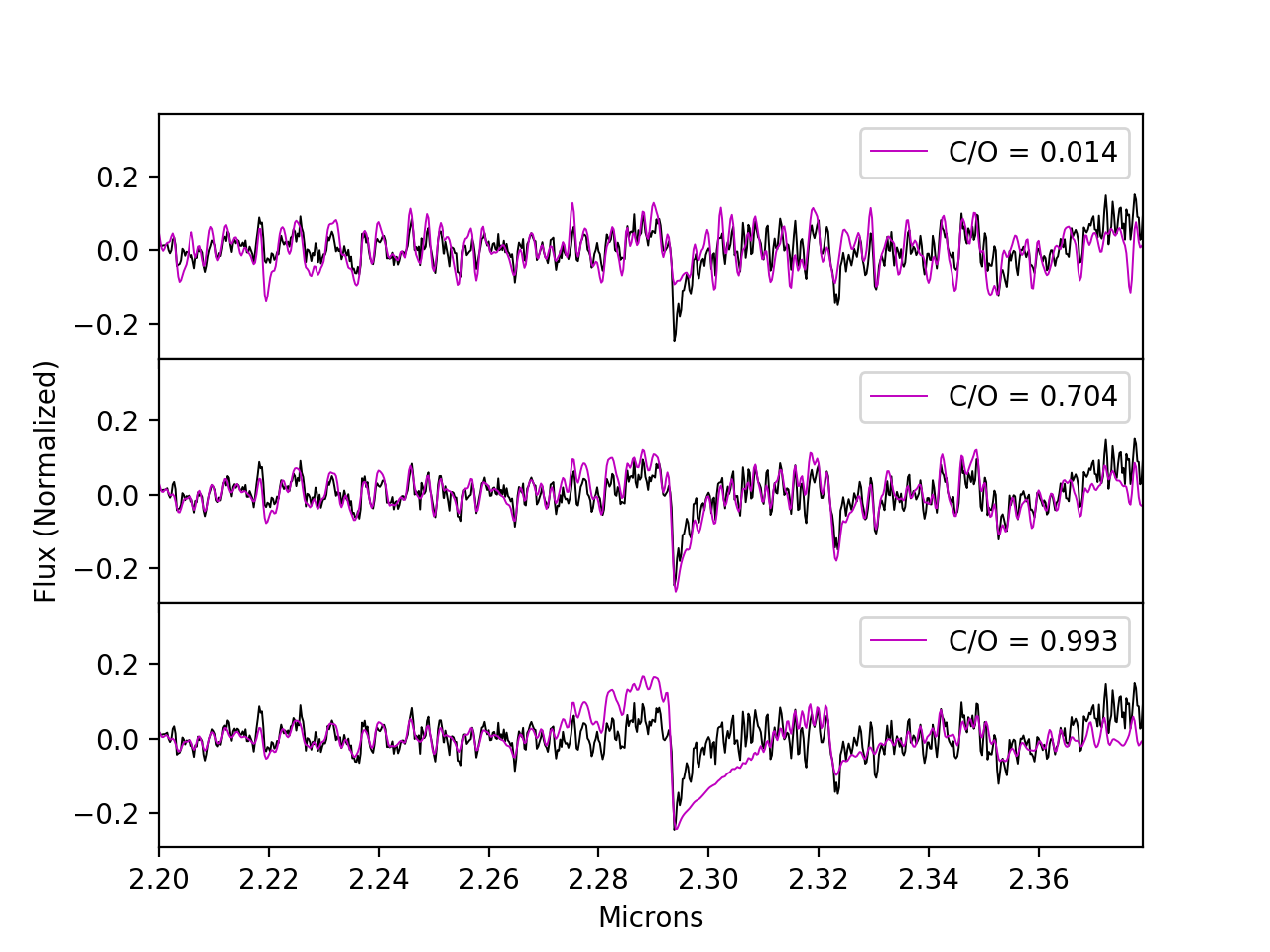}
\caption{Visual comparison of three different $T_\mathrm{eff}$ = 2048 K and $\log g$ = 3.77 models with different values of C/O in our scaled mole fraction grid.  The best fit C/O ratio is shown in the central panel, while values of very high (bottom) and very low (top) C/O ratio are clearly disfavored by our data.  The relative strengthening or weakening of the primary CO bandhead at $\sim$2.29 $\mu$m is a fairly clear discriminator at R$\sim$4000 that might otherwise be lost at lower spectral resolution.}
\label{fig:co_comparison}
\end{figure*}

The C/O ratio dependence on atmospheric mole fractions (N) is

\[\frac{C}{O}=\frac{N(CH_4)+N(CO)}{N(H_2O)+N(CO)},\]

\noindent and for small amounts of CH$_{4}$, as in $\kappa$ And b's case, the C/O ratio can be determined by H$_2$O and CO alone \citep{barman15}. The C/O ratio we derive for $\kappa$ And b is 0.70$_{-0.24}^{+0.09}$. In Figure \ref{fig:co_comparison} we show a visual comparison of three different models with different values of C/O, with our best-fit model in the middle panel. Clearly, the models with low C/O do not make deep enough lines in the CO bandhead, and the models with C/O near unity make the first drop in the CO bandhead too wide. With lower resolution, it would be difficult to distinguish this difference, thus demonstrating the need for higher spectral resolution to probe these abundance ratios.

Due to the remaining uncertainty in the temperature of the planet, we verified that lower temperature model grids with scaled mole fractions return C/O ratios emcompassed by our model.  We explored a grid with a temperature of $\sim$1900 K and a $\log g\sim$4, scaling the ratios of H$_2$O and CO by the same values as above.  We find that the best matching spectrum at this temperature is also $\sim$0.70, with similar uncertainties.  Given the obvious changes in the spectral morphology expected at high and low C/O ratio as shown in Figure \ref{fig:co_comparison}, it is not surprising that a small temperature change does not dramatically change the best-fit C/O ratio.  Thus we assert that our uncertainties properly capture our current knowledge of the C/O ratio for $\kappa$ And b.

\section{Kinematic Modeling}\label{sec:rvs}

\subsection{Radial Velocity Measurement}

Radial velocity measurements can be used to help determine the orientation of the planets' orbital plane. We measure the radial velocity of $\kappa$ And b following a similar method to the one described in \citet{ruffio19}. A significant limitation of \citet{ruffio19} is that the transmission of the atmosphere in \citet{ruffio19} is calculated using A0 star calibrators, which assumes that the tellurics are not changing during the course of a night. This assumption is not valid for the $\kappa$ And b data presented in this work, as discussed in Section \ref{sec:data_red}.  We therefore improved upon the method to correct for the biases due to the variability of the telluric lines compared to the calibrator. 

A common way to address such systematics in high-resolution spectroscopic data is to use a principal component analysis (PCA)-based approach to subtract the correlated residuals in the data \citep{Hoeijmakers2018,PetitditdelaRoche2018,WangJi2018}. However, this approach can lead to over-subtraction of the planet signal and therefore also bias any final estimation. For example, the water lines from the companion can be subtracted by the telluric water lines appearing in the PCA modes. The over-subtraction can be mitigated by jointly fitting for the planet signal and the PCA modes, which is possible in the framework presented in \citet{ruffio19}. The original data model is,
\begin{equation}
     {\bm d} = {\bm M}_1{\bm \phi}_1 + {\bm n}.
\end{equation}
The data ${\bm d}$ is a vector including the pixel values of a spectral cube stamp centered at the location of interest. The data vector has $N_{\bm d}=5\times5\times N_\lambda$ elements corresponding to a $5\times5$ spaxel stamp in the spatial dimensions and $N_\lambda$ spectral channels (e.g., $N_\lambda = 1665$ in $K$-band).
The matrix ${\bm M}_1$ includes a model of the companion and the spurious starlight. It is defined as ${\bm M}_{1}=[{\bm c}_{\mathrm{0,planet}},{\bm c}_1,\dots,{\bm c}_{25}]$, where the ${\bm c}_i$ are column vectors with the same size as the data vector ${\bm d}$. The companion model ${\bm c}_{\mathrm{0,planet}}$ is also a function of the RV of the companion.
The linear parameters of the model are included in ${\bm \phi}_1$ and the noise is represented by the random vector ${\bm n}$.

A spectrum of the planet can be extracted at any location in the image by, first, subtracting a fit of the null hypothesis (i.e., ${\bm M}_{0}=[{\bm c}_1,\dots,{\bm c}_{25}]$) from the data, ${\bm r}_{xy} = {\bm d} - {\bm M}_{0}{\bm \phi}_0$, and then, fitting the companion PSF at each spectral channel to the residual stamp spectral cube. 
We perform this operation at each location in the field of view and divide the subsequent residual spectra by their local low-pass filtered data spectrum.

This results in a residual vector ${\bm r}_{xy}$, which has been normalized to the continuum, for each spaxel in the field of view. After masking the spaxels surrounding the true position of the companion, a PCA of all the ${\bm r}_{xy}$ for a given exposure defines a basis of the residual systematics in the data. These principal components can be used to correct the model of the data.
Before they can be included in the data model, each principal component needs to be rescaled to the local continuum, which is done by multiplying them by the low-pass filtered data at the location of interest. Finally, these 1D spectra are applied to the 3D PSF to provide column vectors that can be used in the model matrix ${\bm M}$.
We denote these column vectors $\{ {\bm r}_{\mathrm{pc}1},{\bm r}_{\mathrm{pc}2},\dots\}$ ordered by decreasing eigenvalues.

A new data model ${\bm M}_2$ including the first $K$ principal components is defined as
\begin{equation}
    {\bm M}_2 = [{\bm c}_{\mathrm{0,planet}},{\bm c}_1,\dots,{\bm c}_{25},{\bm r}_{\mathrm{pc}1},\dots,{\bm r}_{\mathrm{pc}K}].
    \label{eq:modelwithPCA}
\end{equation}
We define a new vector of linear parameter ${\bm \phi}_2$ including $K$ more elements than ${\bm \phi}_1$.
The advantage of this approach is that the PCA modes are jointly fit with the star and the companion models preventing over-subtraction. Additionally, the general form of the linear model is unchanged, which implies that the radial velocity estimation is otherwise identical to \citet{ruffio19}.

\begin{figure*}
  \centering
  \includegraphics[width=1\linewidth]{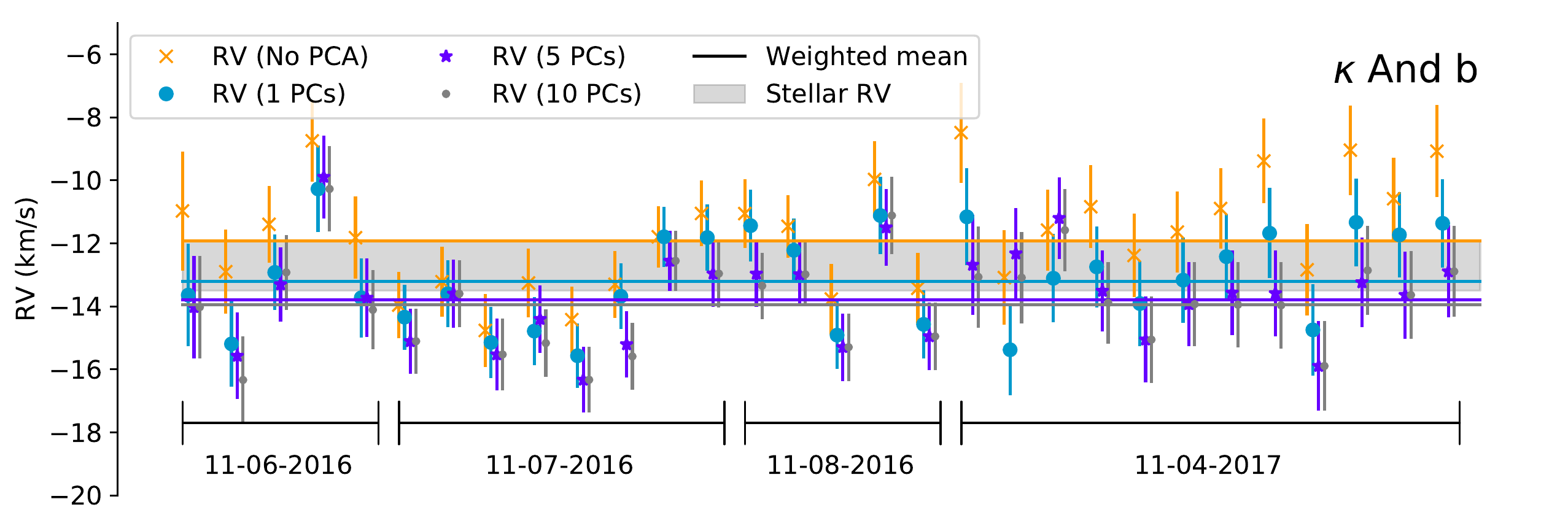}
  \caption{Radial velocity (RV) measurements of $\kappa$ And b by individual exposures and epoch of observation. The grey region represents the current uncertainty in the RV of the star. The RVs are shown for different number of principal components (None, 1, 5, and 10 respectively) included in the data model. The weighted mean RVs (solid horizontal lines) converge as the number of principal components increases. The final RV values and uncertainties are available in Table 3.}
  \label{fig:kapAndbRV}
\end{figure*}

Figure \ref{fig:kapAndbRV} shows the RV estimates for each exposure as a function of the number of principal components used in the model. The final RV converges from  $-11.9\pm0.4\,\mathrm{km}\,\mathrm{s}^{-1}$ to $-13.9\pm0.4\,\mathrm{km}\,\mathrm{s}^{-1}$ as the number of modes increases suggesting a  $2\,\mathrm{km}\,\mathrm{s}^{-1}$ bias in the original model.  In order to increase our confidence in the robustness of the RV estimate and uncertainty, we calculate the final RV and uncertainty after binning the data by pair to account for possible correlations between exposures. Each pair of measurements is replaced by their mean value and largest uncertainty. We note that the reduced $\chi^2$ is lower than unity, which suggests that the final uncertainty is not overestimated.

\begin{figure*}
  \centering
  \includegraphics[width=1\linewidth]{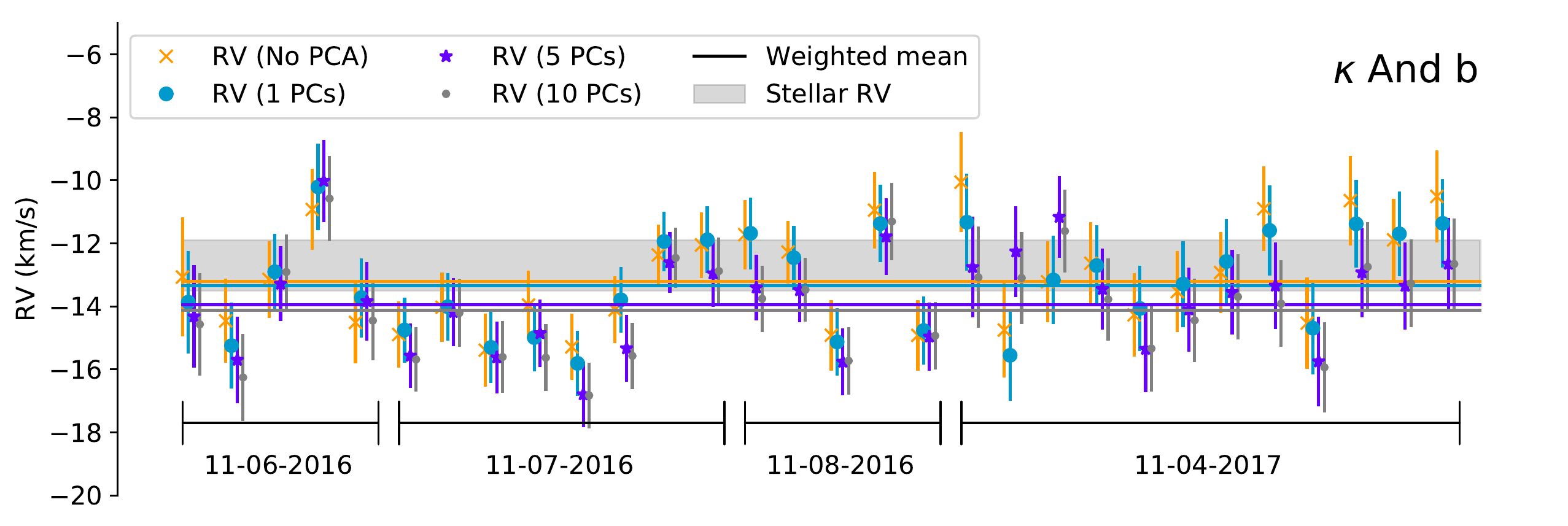}
  \caption{Same as Figure \ref{fig:kapAndbRV}, but corrected for biases using simulated planet injection and recovery. The final RV values and uncertainties are available in Figure 3.}
  \label{fig:kapAndbRV_fakes_corrected}
\end{figure*}

Additionally, we perform a simulated companion injection and recovery at each location in the field of view to estimate possible residual biases in the data. The corrected RV estimates are shown in Figure \ref{fig:kapAndbRV_fakes_corrected}, which prove to be consistent with the results from Figure \ref{fig:kapAndbRV}. Table 3 summarizes the RV estimates, uncertainties, and $\chi_r^2$ as a function of the different cases presented previously. The uncertainties are inflated when $\chi_r^2$ is greater than unity.

\begin{deluxetable*}{@{\extracolsep{4pt}}c|lr|lr|lrr|lrr} 
\tablewidth{0pc}
\tabletypesize{\scriptsize}
\label{tab:RVs}
\tablecaption{$\kappa$ And b RV estimates summary.} 
\tablehead{ 
\colhead{}&
\multicolumn{2}{c}{Independent}&
\multicolumn{2}{c}{Binned}&
\multicolumn{3}{c}{Independent + injection \& recovery}&
\multicolumn{3}{c}{Binned + injection \& recovery} \\
\cline{2-3} \cline{4-5} \cline{6-8} \cline{9-11}
  \colhead{\# PCs} & 
  \colhead{RV} & \colhead{$\chi^2_r$} & 
  \colhead{RV} & \colhead{$\chi^2_r$} & 
  \colhead{RV} & \colhead{$\chi^2_r$} & \colhead{Offset} &
  \colhead{RV} & \colhead{$\chi^2_r$} & \colhead{Offset} \\
  \colhead{} & 
  \colhead{($\mathrm{km}\,\mathrm{s}^{-1}$)} & \colhead{} & 
  \colhead{($\mathrm{km}\,\mathrm{s}^{-1}$)} & \colhead{} & 
  \colhead{($\mathrm{km}\,\mathrm{s}^{-1}$)} & \colhead{} & \colhead{($\mathrm{km}\,\mathrm{s}^{-1}$)} &
  \colhead{($\mathrm{km}\,\mathrm{s}^{-1}$)} & \colhead{}& \colhead{($\mathrm{km}\,\mathrm{s}^{-1}$)}
}
\startdata 
None & $-11.9\pm0.4\tablenotemark{a}$ & $1.8$ & 
       $-11.9\pm0.3\tablenotemark{a}$ & $1.1$ &
       $-13.2\pm0.3\tablenotemark{a}$ & $1.6$ & $-1.28$ & 
       $-13.1\pm0.3$ & $0.8$ & $-1.28$  \\
1    &  $-13.2\pm0.3\tablenotemark{a}$ & $1.5$ & 
       $-13.1\pm0.3$ & $0.8$ &
       $-13.3\pm0.4\tablenotemark{a}$ & $1.6$ & $-0.14$ & 
       $-13.3\pm0.3$ & $0.9$ & $-0.15$  \\
5    & $-13.8\pm0.3\tablenotemark{a}$ & $1.4$ & 
       $-13.8\pm0.3$ & $0.8$ &
       $-14.0\pm0.3\tablenotemark{a}$ & $1.6$ & $-0.16$ & 
       $-14.0\pm0.3$ & $0.9$ & $-0.17$  \\
10    & $-13.9\pm0.3\tablenotemark{a}$ & $1.4$ & 
       $-13.9\pm0.3$ & $0.8$ &
       $-14.1\pm0.4\tablenotemark{a}$ & $1.6$ & $-0.19$ & 
       $-14.1\pm0.3$ & $1$ & $-0.19$
\enddata
\tablenotetext{-}{(Columns 2--3) RVs calculated using a data model that includes principal component as defined in Equation \ref{eq:modelwithPCA}. The final RVs were calculated with a weighted mean assuming that each individual exposure is independent. (Columns 4--5) Same as columns 2--3, but pairs of consecutive exposures were averaged and the largest of their uncertainties used. (Columns 6--7) RVs are corrected for biases using simulated planet injection and recovery. The resulting offset on the final RV with and without the injection and recovery is given in column 8. (Columns 9--11) Same as columns 6--8, but combining consecutive pairs of exposures.}
\tablenotetext{a}{Uncertainties have been inflated by $\chi^2_r$ when $\chi^2_r$ is greater than unity.}
\end{deluxetable*}

We conclude that the RV of $\kappa$ And b is $-14.1\pm0.4\,\mathrm{km}\,\mathrm{s}^{-1}$ (cf Table 4), while the estimates for the RV of the star are $-12.7\pm0.8\,\mathrm{km}\,\mathrm{s}^{-1}$ \citep{Gontcharov2006} and $-11.87\pm1.53\,\mathrm{km}\,\mathrm{s}^{-1}$ \citep{Becker2015}. These values are consistent within the uncertainties - we use $-12.7\pm0.8\,\mathrm{km}\,\mathrm{s}^{-1}$ in the following because the uncertainty is smaller. The relative RV between the companion and the star is $-1.4\pm0.9\,\mathrm{km}\,\mathrm{s}^{-1}$ for which the error is dominated by the stellar RV. Similar to \citet{ruffio19}, this highlights the need to better constrain the stellar RV of stars hosting directly imaged companions.

\begin{deluxetable}{lc} 
\tabletypesize{\scriptsize} 
\tablewidth{0pt} 
%\rotate
\tablecaption{Final RVs for $\kappa$ Andromedae b.} 
\label{tab:RVsummary}
\tablehead{ 
  \colhead{Date} & \colhead{RV ($\mathrm{km}\,\mathrm{s}^{-1}$)}
}
\startdata 
2016 Nov 6--8 & $-14.3\pm0.4$  \\
2017 Nov 4 & $-13.6\pm0.6$   
\enddata
\end{deluxetable}
\pagebreak

\subsection{Orbital Analysis}

The orbit of $\kappa$ And b has been explored using astrometry by several authors (\citealt{Blunt17,currie18,uyama20,Bowler19}).  Though the orbit is highly under-constrained in terms of phase coverage, current fits to astrometry have yielded some constraints on orbit orientation and eccentricity of the companion.  In particular, the eccentricity is currently estimated to be fairly high ($>$0.7).  

The measurement of an RV for the companion with our OSIRIS data offers a valuable new piece of information, wherein degeneracies in the orbit orientation can be resolved.  To determine the constraints provided by the RV measurement, we performed a series of orbit fits with both astrometry from the literature (\citealt{Carson13,Bonnefoy14,currie18,uyama20}) and our OSIRIS RV using the code described in \citet{Kosmo-Oneil19}.  Specifically, we use the Efit5 code \citep{meyer12}, which uses MULTINEST to perform a Bayesian analysis of our data (e.g., \citealt{feroz09}), and we use two different priors.  We first use the typical flat priors in orbital parameters, including period (P), eccentricity (e), time of periastron passage (T$_0$), inclination (flat in $\sin i$), longitude of the ascending node ($\Omega$ or O), and longitude of periastron passage ($\omega$ or w).  We also use the observational-based priors derived in \citet{Kosmo-Oneil19}.  Although we believe the latter are more appropriate in this case due to the biases introduced by flat priors for under-constrained orbits, we include both for completeness.  We performed fits both with and without the RV point derived above to determine the impact of including the RV. We fix the distance to 50.0 $\pm$ 0.1 pc \citep{gaia18}, and the mass to the value of 2.7 $\pm$ 0.1 M$_{\odot}$ estimated by \citet{jones16}, which encompasses the range of values they found given uncertainty in the internal metallicity of $\kappa$ And A.

\begin{figure*}
  \centering
  \includegraphics[width=1\linewidth]{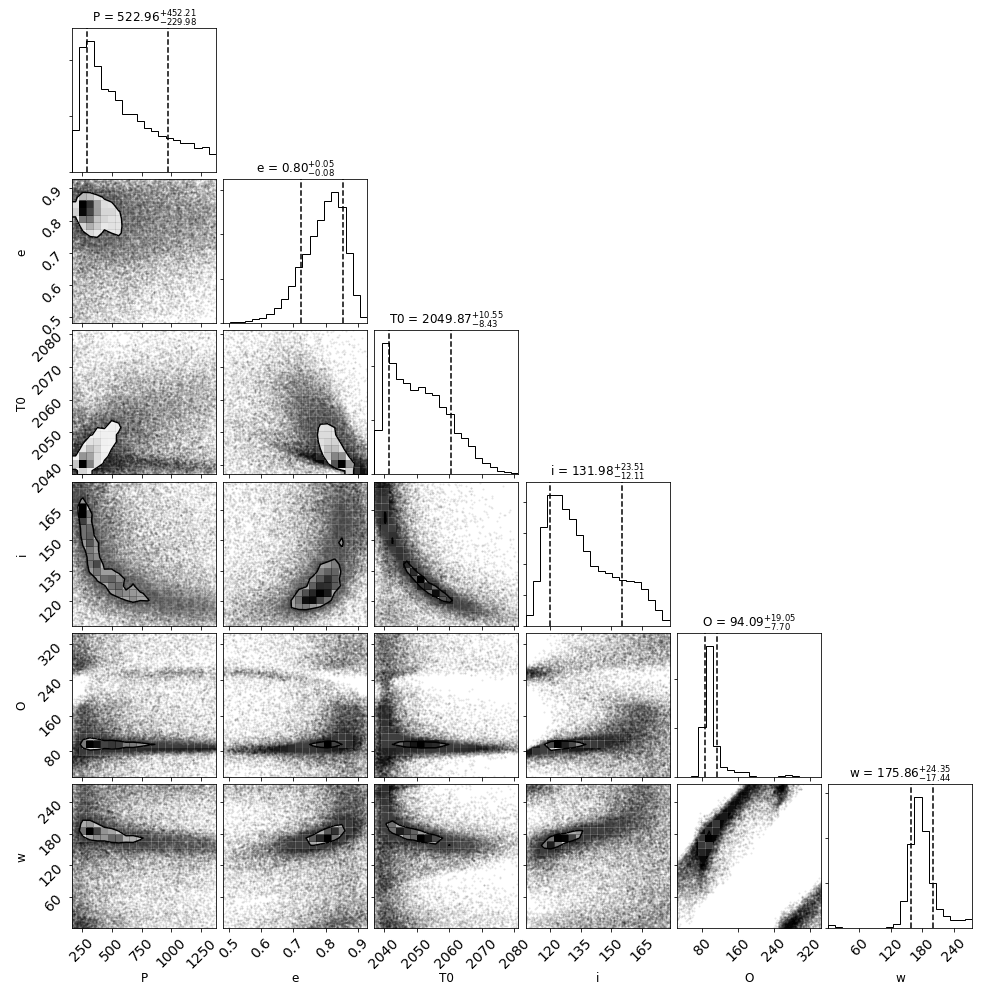}
  \caption{Corner plot showing the results of fitting the orbit of $\kappa$ And b, including both astrometry and radial velocities.  In this case, we use the observationally-based prior presented in \citet{Kosmo-Oneil19}, which can help account for biases in parameters like T$_0$ that arise in undersampled orbits.  Note that this prior does increase the range of T$_o$ included in our 1$\sigma$ uncertainties more than is seen when flat priors in the orbital parameters are used (e.g., Figure \ref{fig:corner_flat_rv}).  The allowed parameter space for $\omega$ also shrinks considerably when RV is included (see for comparison Figure \ref{fig:corner_obs_no_rv}).}
  \label{fig:corner_obs_rv}
\end{figure*}

\begin{figure*}
  \centering
  \includegraphics[width=1\linewidth]{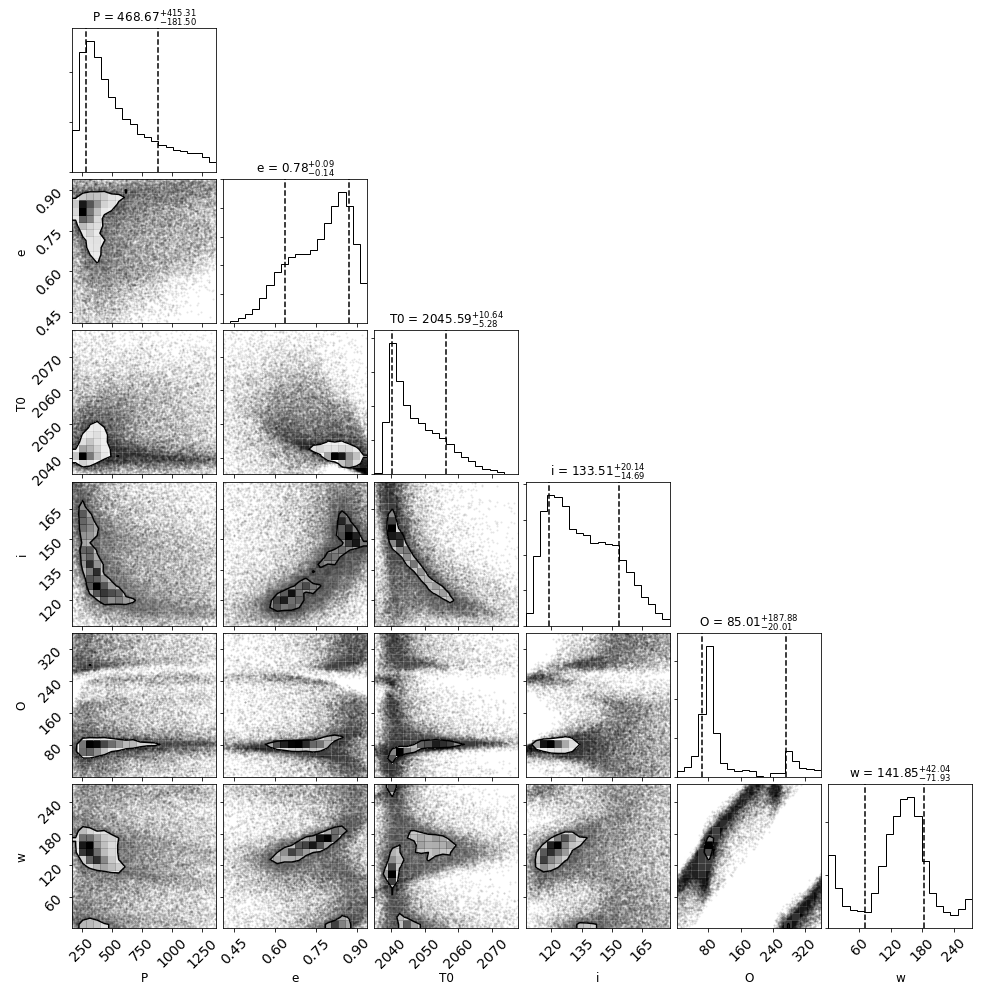}
  \caption{The same as Figure \ref{fig:corner_obs_rv}, but with no RV included in the fit.  The results give values for orbital parameters consistent with previous fits found in the literature.  A secondary peak in $\Omega$ can be seen more prominently here around $\sim$270$^o$ that is nearly absent in Figure \ref{fig:corner_obs_rv}.  The addition of the RV eliminates this degeneracy in the orbit plane orientation.}
  \label{fig:corner_obs_no_rv}
\end{figure*}

\begin{figure*}
  \centering
  \includegraphics[width=1\linewidth]{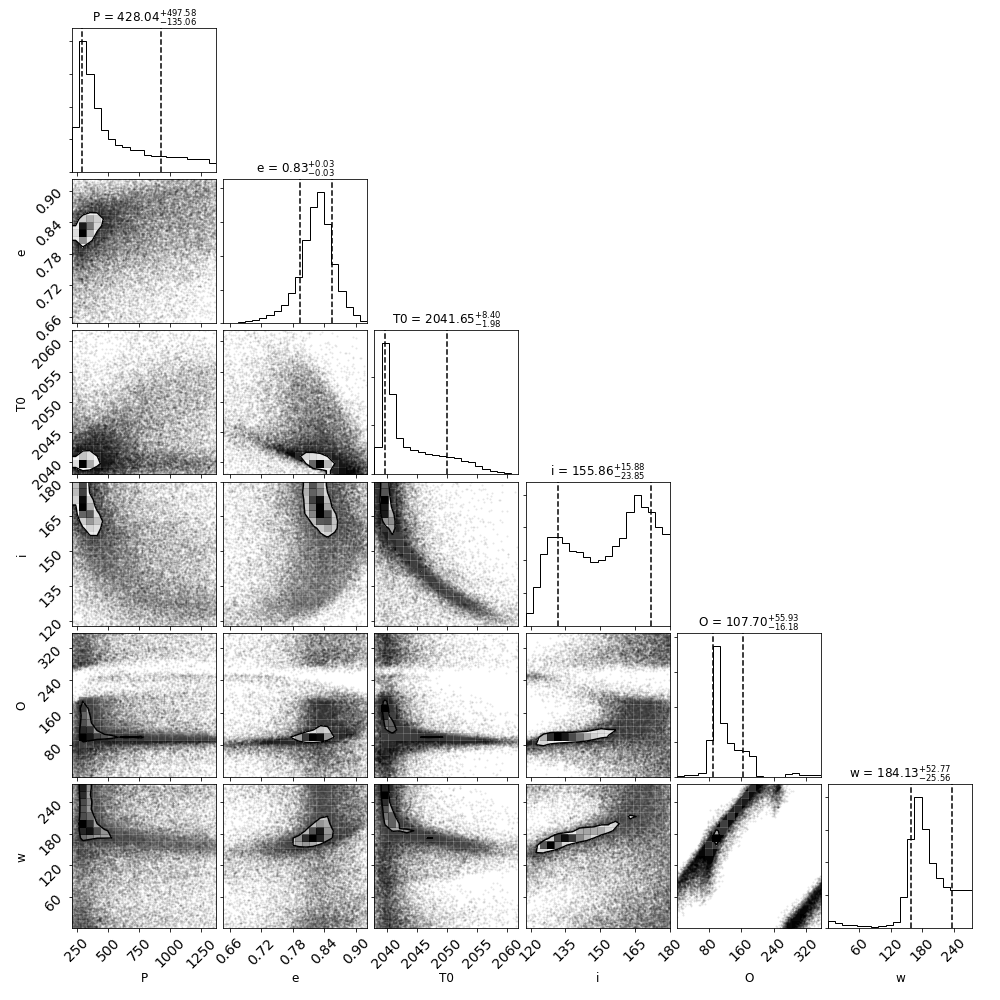}
  \caption{Corner plot showing the results of fitting the orbit of $\kappa$ And b, including both astrometry and radial velocities.  In this case, we use the typical flat priors in fit parameter space for easier comparison to previous work.  Note that the use of these priors leads to a highly peaked prediction for T$_0$.  Since velocity changes rapidly at this orbital phase, we will be able to test whether this prediction holds true in the next few years (Figure \ref{fig:orbit_model_flat}).}
  \label{fig:corner_flat_rv}
\end{figure*}

\begin{figure*}
  \centering
  \includegraphics[width=1\linewidth]{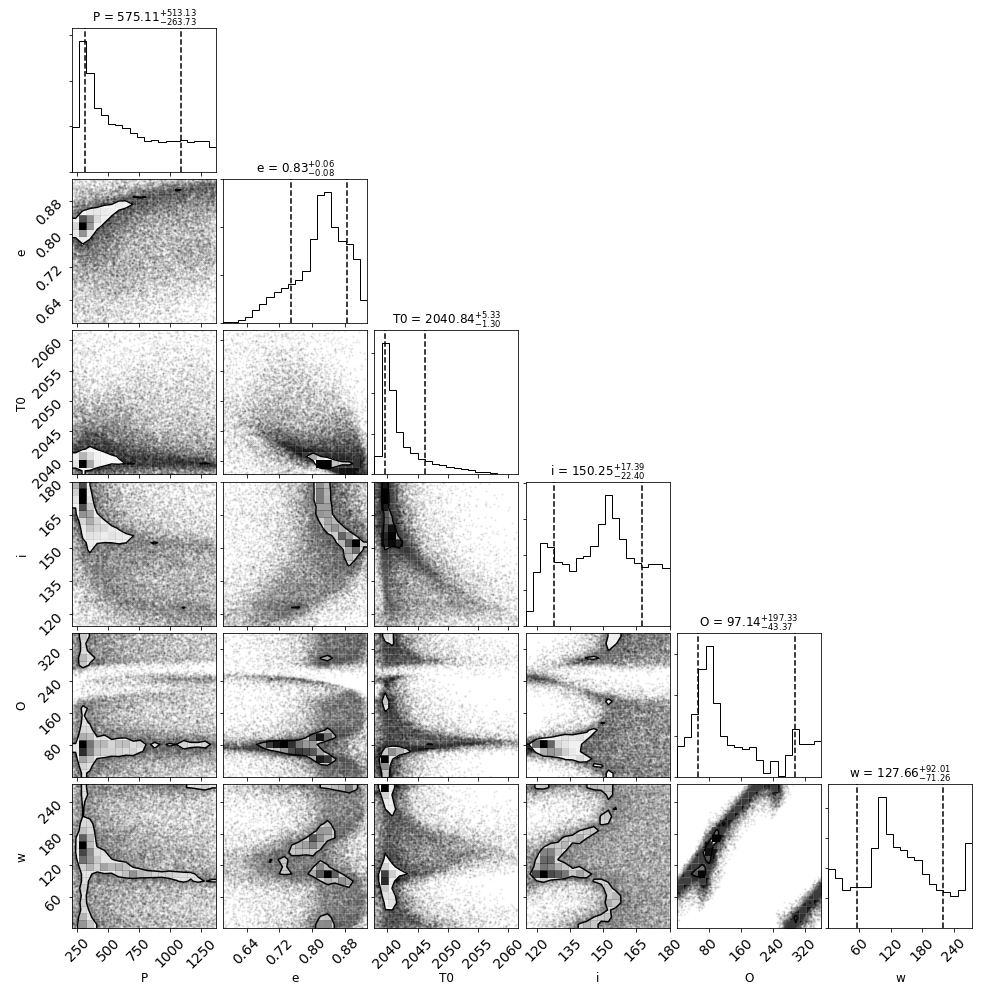}
  \caption{The same as Figure \ref{fig:corner_flat_rv}, but with no RV included in the fit.  The results give values for orbital parameters consistent with previous fits found in the literature.  Again, a secondary peak in $\Omega$ can be seen more prominently here around $\sim$270$^o$ that is nearly absent in Figure \ref{fig:corner_flat_rv}.  The addition of the RV eliminates this degeneracy in the orbit plane orientation (regardless of prior choice).}
  \label{fig:corner_flat_no_rv}
\end{figure*}

\begin{figure*}
  \centering
  \includegraphics[width=1\linewidth]{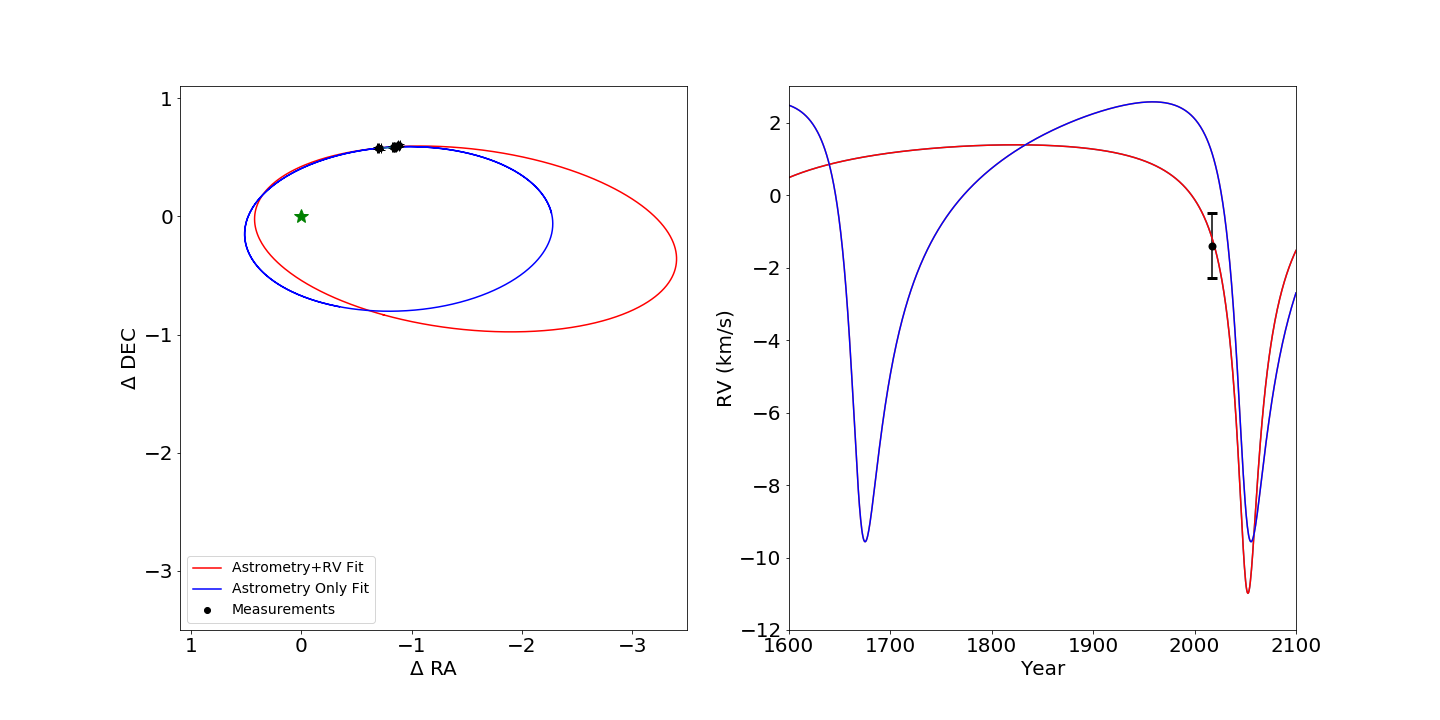}
  \caption{Best-fit orbits with (red) or without (blue) inclusion of the OSIRIS RV using the observationally-based prior.  Because of the large parameter space allowed by the astrometry, the best-fits are used here only for illustrative purposes.  The left panel shows the orbits on the plane of the sky while the right panel demonstrates the variation in relative RV of the planet with time.  Including the RV increases the preferred eccentricity of the best-fit solution, though the astrometry drives solutions to high eccentricities regardless.  Based on the right hand panel, the RVs clearly have more diagnostic power in the next several years than the astrometry.  If the planet is indeed approaching periastron passage, a rapid decrease in the relative radial velocity is predicted.}
  \label{fig:orbit_model_obs}
\end{figure*}

\begin{figure*}
  \centering
  \includegraphics[width=1\linewidth]{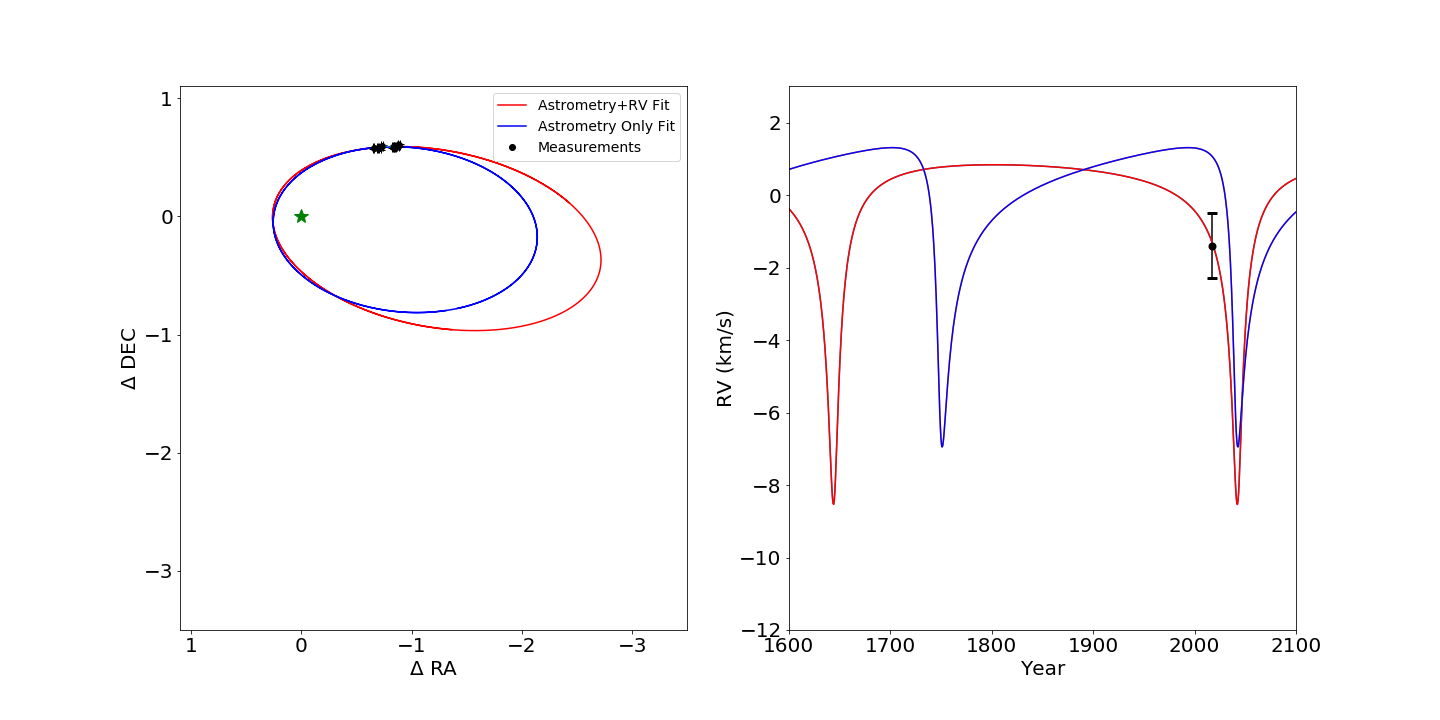}
  \caption{The same as Figure \ref{fig:orbit_model_obs}, but using flat orbit parameter priors.  Either choice of prior yields preferred orbit solutions that approach periastron in the next several years, stressing the utility of more RVs.}
  \label{fig:orbit_model_flat}
\end{figure*}

The results of these fits are given in Table \ref{tab:orbit} and shown visually in corner plots in Figures \ref{fig:corner_obs_rv}--\ref{fig:corner_flat_no_rv}.  The addition of the RV constrains $\Omega$ to most likely be $\sim$85--90$\deg$, although due to the large uncertainty in the RV the secondary peak is not completely ruled out.  Additionally, the RV pushes the distribution of eccentricities slightly higher than the astrometry alone, with global minima $>$0.8, although the uncertainties encompass the previous values.  Figures \ref{fig:orbit_model_obs} and \ref{fig:orbit_model_flat} demonstrates the impact of the RV on the best-fit orbits.  Although the best fits are not strictly meaningful due to undersampling of the period, there are clear differences in orbit predictions when the RV is included - the best fit with astrometry alone favors RVs that are closer to 0 km s$^{-1}$.  

The current prediction (whether RVs are included or not) is that $\kappa$ And b is on its way towards closest approach in the next 20--30 years.  It is possible this prediction is impacted by systematics in the astrometric dataset, which is drawn from multiple different cameras and reduction pipelines.  Indeed, the observational prior is meant to account for this known bias in T$_{0)}$, and using it pushes the prediction of periastron later by about 10 years (Table \ref{tab:orbit}).  If it is the case that the planet is heading towards closest approach, the predicted change in RV in the next several years is significant and thus can be easily confirmed with more data of similar quality in the next decade.  Thus spectroscopy has the potential to provide much more stringent constraints on the orbit in the near term than more astrometric measurements.

%\floattable
\begin{turnpage}
\begin{deluxetable}{@{\extracolsep{1pt}}l|cccccc|cccccc|cccccc} 
\tabletypesize{\footnotesize} 
\tablewidth{0pt} 
\tablecaption{Derived orbit parameters for $\kappa$ And B.
\label{tab:orbit}} 
%\rotate
\tablehead{ 
\colhead{}&
\multicolumn{6}{c}{Global Minimum}&
\multicolumn{6}{c}{Mean}&
\multicolumn{6}{c}{1$\sigma$ Range}\\
\cline{2-7} \cline{8-13} \cline{14-19} 
  \colhead{Fit Type} & 
  \colhead{P} & \colhead{ecc} & 
  \colhead{T$_o$} & \colhead{inc.} & 
  \colhead{$\omega$} & \colhead{$\Omega$} &  \colhead{P} & \colhead{ecc} & 
  \colhead{T$_o$} & \colhead{inc.} & 
  \colhead{$\omega$} & \colhead{$\Omega$}&  \colhead{P} & \colhead{ecc} & 
  \colhead{T$_o$} & \colhead{inc.} & 
  \colhead{$\omega$} & \colhead{$\Omega$} \\
  \colhead{} & 
  \colhead{(yrs)} & \colhead{} & 
  \colhead{(yr)} & 
  \colhead{(deg.)} & \colhead{deg.} & \colhead{(deg)} & \colhead{(yrs)} & \colhead{} & 
  \colhead{(yr)} & 
  \colhead{(deg.)} & \colhead{deg.} & \colhead{(deg)} & \colhead{(yrs)} & \colhead{} & 
  \colhead{(yr)} &
  \colhead{(deg.)} & \colhead{deg.} & \colhead{(deg)} 
}
\startdata 
Flat Priors  & 398.0 & 0.83 & 2042.5 & 152.1 & 186.4 & 104.4 & 428.0 & 0.83 & 2041.7 & 155.9 & 184.1 & 107.1 & 293.0 -  & 0.80 - & 2039.7 -  & 132.0 -  & 158.6 -  & 91.5 - \\
Astrometry+RV &  &  &  & & & &  &  &  &  &  &  &  925.6 & 0.86 &  2050.1 & 171.7 &  236.9 & 163.6 \\
\hline
Flat Priors & 291.42 & 0.79 & 2040.85 & 156.2 & 68.5 & 149.6 & 576.1 & 0.83 & 2040.8 & 150.3 & 127.7 & 97.1 & 311.4 - & 0.75 - & 2039.5 - & 127.9 - & 56.4 - & 53.8 - \\
Astrometry Only &  &  &  & & & &  &  &  &  &  &  & 1088.24 & 0.89 &  2046.2 & 167.6 &  219.7 & 294.5 \\
\hline
Obs. Priors & 576.6 & 0.78 & 2051.7 & 129.3 & 172.5 & 92.5 & 523.0 & 0.80 & 2049.9 & 132.0 & 175.9 & 94.1 & 293.0 - & 0.72 - & 2041.4 - & 119.9 - & 158.4 - & 86.4 - \\
Astrometry+RV &  &  &  & & & &  &  &  &  &  &  &  975.2 & 0.85 &  2060.4 & 113.1 &  200.2 & 163.6 \\
\hline
Obs. Priors & 380.4 & 0.66 & 2050.8 & 127.3 & 150.7 & 78.1 & 468.7 & 0.78 & 2045.6 & 133.5 & 141.9 & 85.0 & 287.2 - & 0.64 - & 2040.3 - & 118.8 - & 69.9 - & 65.0 - \\
Astrometry Only &  &  &  & & & &  &  &  &  &  &  & 884.0 & 0.87 &  2056.2 & 153.7 &  183.9 & 272.9 \\
\enddata
\end{deluxetable}
\clearpage
\end{turnpage}

\section{Discussion and Conclusions}\label{sec:discussion}

Using moderate-resolution spectroscopy, we have greatly expanded our knowledge of the low-mass, directly imaged companion, $\kappa$ And b. In recent years, most studies of the $\kappa$ And system have led to the conclusion that it is young, as originally predicted by \cite{Carson13}.  Our derivation of low surface gravity ($\log g<$4.5) using our OSIRIS spectrum is another piece of evidence in favor of a young age.  If we consider the age range adopted by \cite{jones16} of 47$^{+27}_{-40}$ Myr, the predicted mass for a roughly $\sim$2050 K object range from $\sim$10--30 $M_\mathrm{Jup}$ (e.g., \citealt{baraffe15}).  We note that our best-fit surface gravity of 3.8 is too low to be consistent with evolutionary models for this mass range, which predicts $\log g \approx$ 4--4.7.  However, our uncertainties allow for gravities up to log(g)$\sim$4.5.  The OSIRIS data does not favor $\log g$ greater than 4.5, which argues for an age less than $\sim$50 Myr.  Our derived radius is also on the low end of what is allowed by evolutionary models, which predict R = 1.3 R$_{Jup}$ for older, more massive objects through R = 1.8 R$_{Jup}$ for younger, lower mass objects.  Our uncertainties again are sufficient to encompass this range.  The implied bolometric luminosity is consistent with \citet{currie18}, who note that it is similar to other young, substellar objects.

Additional constraints on the temperature,  cloud properties, and radius in the future via additional photometry, spectra, or modeling could yield tighter constraints on the mass of $\kappa$ And b.  $\kappa$ And b is an excellent candidate for moderate-resolution spectroscopy at shorter wavelengths to look for lines from higher atomic number species beyond carbon and oxygen. With future measurements of highly gravity sensitive lines, like potassium in the $J$-band if detectable, stronger limits can be placed spectroscopically on $\log g$, which will provide a more robust age.  Further mass constrains could also come from astrometric measurements with $Gaia$ or more radial velocity measurements that include velocity measurements for the star, although the precision of such RVs may be limited.

Given the size and separation of $\kappa$ And b, its formation pathway is of considerable interest.  Our measurement of C/O here provides one possible diagnostic of formation.  We note that in a number of recent works, it has been demonstrated that the C/O ratio is impacted by a variety of phenomena beyond formation location in the disk.  These include the grain size distribution \citep{piso15}, migration of grains or pebbles \citep{booth17}, migration of planets themselves \citep{cridland20}, and whether the accreted material is from the midplane (\citealt{morbidelli14,batygin18}).  Current studies are therefore incorporating more chemical and physical processes into models to get a better idea of what impacts the C/O ratio and what exactly the ratio tells us about formation.

With these studies in mind, we turn to the C/O ratio we have measured for $\kappa$ And b.  Although our current uncertainties allow for somewhat elevated C/O ratios, the most likely scenario is that C/O is roughly consistent with the Sun.  This result diagnostically points to a very rapid formation process, potentially through either gravitational instability or common gravitational collapse similar to a binary star system.  The complication, however, is that the comparison must be made to the host star in order to draw definitive conclusions about formation. The C/O ratio of the host star, $\kappa$ And A, has not been measured or reported in the literature.  As a late B-type star, probing these abundances is challenging, although certainly possible \citep{takeda16}.  However, the rapid rotation of $\kappa$ And A ($\sim$162 km s$^{-1}$; \citealt{royer07}), may make abundance determinations difficult.  High resolution optical spectroscopy for the star would be able to probe potential diagnostic lines, such as the OI triplet at 7771 \AA.  Until individual abundance estimates for C and O are available, however, we can only conclude that the evidence points to roughly similar values for the host star and the companion if the star has similar abundances to the Sun.

In terms of overall metallicity, the [Fe/H] abundance of $\kappa$ And A was estimated by \cite{wu11} to be subsolar, [M/H] = $-0.32 \pm$ 0.15.  However, \citet{jones16} argue this is unlikely to be the true internal metallicity of the star, instead adopting a roughly solar abundance range of [M/H] = 0.00$\pm$0.14 based on the range of metallicities in nearby open clusters.  Interestingly, our slightly subsolar best-fit metallicity for $\kappa$ And b may suggest that indeed the star is metal poor overall.  A number of theoretical works have suggested that formation via gravitational instability would preferentially occur around low metallicity stars.  Metal poor gas allows for shorter cooling timescales, allowing planets to quickly acquire sufficient density to avoid sheering (e.g., \citealt{boss02,cai06,helled11}). Since metals are difficult to measure in high mass hosts like $\kappa$ And A, direct metallicity measurements of the planets themselves could provide insight into measurements for the host star.  We note that the derived abundances could be impacted by non-equilibrium chemistry effects in the $K$-band, and measuring atomic abundances can mitigate this issue and may be preferable (e.g., \citealt{nikolov18}). Additional metallicity measurements for directly imaged planets will also help probe the intriguing trend that the correlation of planet occurrence and metallicity breaks down at $\sim$4 M$_\mathrm{Jup}$ \citep{santos17}.  The apparently low metallicity of $\kappa$ And b is certainly consistent with this finding.

$\kappa$ And b now represents a fourth case of a directly imaged planet, in addition to three of the the HR 8799 planets \citep{Konopacky13,barman15,molliere20}, where the C/O ratio formation diagnostic did not reveal ratios that clearly point to formation via core/pebble accretion.  The scenario certainly cannot be ruled out given the uncertainties in the data and the range of possible C/O ratios predicted by models (e.g., \citealt{madhu19}).  Because of this uncertainty, other probes of formation will be needed to shed additional light on this fascinating population of companions.  That includes the suggestion that the the high eccentricity of $\kappa$ And b is a result of scattering with another planetary mass object.  Our results cannot shed light on potential formation closer to the star using C/O as a diagnostic until we can improve our uncertainties.  Since the C/O ratio is largely a function of the amount of solids incorporated into the atmosphere, it is possible that the massive size of these planets simply implies that they very efficiently and rapidly accreted their envelopes.  This could have included enough solid pollution in the envelope to return the C/O ratio to the original value.  Indeed, there are pebble accretion scenarios proposed in which it is possible to achieve slightly superstellar C/H and C/O, but stellar O/H ratios via significant accretion of large, metal-rich grains \citep{booth17}, which is consistent with our results for $\kappa$ And b.

The next steps for the $\kappa$ And system going forward will be confirmation of the high eccentricity solutions currently favored using more RVs, and continued monitoring with astrometry using consistent instrumentation to limit astrometric systematics.  The strong CO lines and favorable contrast make $\kappa$ And b an excellent candidate for high-resolution, AO-fed spectroscopy with instruments like KPIC on Keck, IRD on Subaru, or CRIRES on the VLT (e.g., \citealt{snellen14,WangJi2018}).  We can also determine whether the bulk population of directly imaged planets show C/O ratios consistent with solar/stellar values by continuing to obtain moderate or high-resolution spectra of these companions.  If the population of directly imaged planets shows C/O distinct from what has been seen with closer in giant planets probed via transmission spectroscopy, this could point to distinct formation pathways for these sets of objects.

\acknowledgements

We would like to thank the referee, Joe Carson, for reviewing this work, and Marshall Perrin and Justin Otor for helpful conversations relating to this work. K. K. W. would also like to thank Thea Kozakis and Laura Stevens for discovering $\kappa$ And b. J.-B. R. acknowledges support from the David \& Ellen Lee Prize Postdoctoral Fellowship. Work conducted by Laci Brock and Travis Barman was supported by the National Science Foundation under Award No. 1405504. Support for this work was provided by NASA through the NASA Hubble Fellowship grant HST-HF2-51447.001-A awarded by the Space Telescope Science Institute, which is operated by the Association of Universities for Research in Astronomy, Inc., for NASA, under contract NAS5-26555. Material presented in this work is supported by the National Aeronautics and Space Administration under Grants/Contracts/Agreements No.NNX17AB63G and NNX15AD95G issued through the Astrophysics Division of the Science Mission Directorate. Any opinions, findings, and conclusions or recommendations expressed in this poster are those of the author(s) and do not necessarily reflect the views of the National Aeronautics and Space Administration. The data presented herein were obtained at the W. M. Keck Observatory, which is operated as a scientific partnership among the California Institute of Technology, the University of California and the National Aeronautics and Space Administration. The Observatory was made possible by the generous financial support of the W. M. Keck Foundation. The authors wish to recognize and acknowledge the very significant cultural role and reverence that the summit of Maunakea has always had within the indigenous Hawaiian community. We are most fortunate to have the opportunity to conduct observations from this mountain.


\begin{thebibliography}{}
\expandafter\ifx\csname natexlab\endcsname\relax\def\natexlab#1{#1}\fi
\expandafter\ifx\csname url\endcsname\relax
  \def\url#1{\texttt{#1}}\fi
\expandafter\ifx\csname urlprefix\endcsname\relax\def\urlprefix{URL }\fi
\providecommand{\eprint}[2][]{\url{#2}}

\bibitem[Allard et al.(2001)]{allard01} Allard, F., Hauschildt, P.~H., Alexander, D.~R., Tamanai, A., \& Schweitzer, A.\ 2001, \apj, 556, 357

\bibitem[Allard et al.(2012)]{allard12} Allard, F., Homeier, D., \& Freitag, B.\ 2012, Philosophical Transactions of the Royal Society of London Series A, 370, 2765


\bibitem[Baraffe et al.(2008)]{Baraffe08} Baraffe, I., Chabrier, G., \& Barman, T., 2008, \aap, 482, 315

\bibitem[Baraffe et al.(2015)]{baraffe15} Baraffe, I., Homeier, D., Allard, F., et al.\ 2015, \aap, 577, A42


\bibitem[Barman et al.(2011)]{Barman11} Barman, T.~S., Macintosh, B., Konopacky, Q.~M., \& Marois, C.\ 2011, \apj, 733, 65

\bibitem[Barman et al.(2015)]{barman15} Barman, T.~S., Konopacky, Q.~M., Macintosh, B., \& Marois, C.\ 2015, \apj, 804, 61

\bibitem[Batygin(2018)]{batygin18} Batygin, K.\ 2018, \aj, 155, 178

\bibitem[Becker et al.(2015)]{Becker2015} Becker, J.~C., Johnson, J.~A., Vanderburg, A., Morton, T.~D.\ 2015, \apj, 217, 29

\bibitem[Bell et al.(2015)]{bell15} Bell, C.~P.~M., Mamajek, E.~E., \& Naylor, T.\ 2015, \mnras, 454, 593


\bibitem[Blake et al.(2010)]{Blake2010} Blake, C.~H., Charbonneau, D., \& White, R.~J.\ 2010, \apj, 723, 684 

\bibitem[Blunt et al.(2017)]{Blunt17} Blunt, S., Nielsen, E.~L, De Rosa, R.~J., et al.\ 2017, \aj, 153, 229 

\bibitem[Bonnefoy et al.(2014)]{Bonnefoy14} Bonnefoy, M., Currie, T., Marleau, G.-D., et al.\ 2014, \aap, 562, A111 

\bibitem[Booth et al.(2017)]{booth17} Booth, R.~A., Clarke, C.~J, Madhusudhan, N., \& Ilee, J.~D.\ 2017, \mnras, 469, 3994 

\bibitem[Boss(2002)]{boss02} Boss, A.~P.\ 2002, \apjl, 567, L149

\bibitem[Bowler(2016)]{Bowler16} Bowler, B.~P.\ 2016, \pasp, 128, 102001 

\bibitem[Bowler et al.(2019)]{Bowler19} Bowler, B.~P., Blunt, S.~C., \& Nielsen, E.~L.\ 2019, arXiv:1911.10569

\bibitem[Brandt, \& Huang(2015)]{brandt15} Brandt, T.~D., \& Huang, C.~X.\ 2015, \apj, 807, 58

\bibitem[Burgasser(2014)]{Burgasser14} Burgasser, A.~J.\ 2014, Astronomical Society of India Conference Series, 7

\bibitem[Burgasser et al.(2016)]{Burgasser16} Burgasser, A.~J., Blake, C.~H., Gelino, C.~R., et al.\ 2016, \apj, 827, 25

\bibitem[Cai et al.(2006)]{cai06} Cai, K., Durisen, R.~H., Michael, S., et al.\ 2006, \apjl, 642, L173


\bibitem[Carson et al.(2013)]{Carson13} Carson, J., Thalmann, C., Janson, M., et al.\ 2013, \apjl, 763, L32 

\bibitem[Chabrier et al.(2000)]{Chabrier00} Chabrier, G., Baraffe, I., Allard, F., \& Hauschildt, P.\ 2000, \apj, 542, 464 

\bibitem[Cridland et al.(2020)]{cridland20} Cridland, A.~J., Bosman, A.~D., \& van Dishoeck, E.~F.\ 2020, \aap, 635, A68

\bibitem[Currie et al.(2018)]{currie18} Currie, T., Brandt, T.~D., Uyama, T., et al.\ 2018, \aj, 156, 291 

\bibitem[David, \& Hillenbrand(2015)]{david15} David, T.~J., \& Hillenbrand, L.~A.\ 2015, \apj, 804, 146

\bibitem[Dodson-Robinson et al.(2009)]{Dodson-Robinson09} Dodson-Robinson, S.~E., Veras, D., Ford, E.B., \& Beichman, C.~A. 2009, \apj, 707, 79 

\bibitem[Feroz (2009)]{feroz09} Feroz, F., Hobson, M.~P., Bridges, M. 2009, \mnras, 398, 1601 


\bibitem[Fitzpatrick \& Massa(2005)]{fitzpatrick05} Fitzpatrick, E.~L., \& Massa, D.\ 2005, \aj, 129, 1642 

\bibitem[Foreman-Mackey et al.(2013)]{ForemanMackey13} Foreman-Mackey, D., Hogg, D. W., Lang, D., \& Goodman, J. 2013, PASP, 125, 306

\bibitem[Gaia Collaboration et al.(2018)]{gaia18} Gaia Collaboration, Brown, A.~G.~A., Vallenari, A., et al.\ 2018, \aap, 616, A1

\bibitem[Gontcharov et al.(2006)]{Gontcharov2006} Gontcharov, G.~A.\ 2006, Astronomy Letters, 32, 759

\bibitem[Goodman \& Weare (2010)]{GoodmanWeare10} Goodman, J., \& Weare, J.\ 2010, Communications in Applied Mathematics and Computational Science, Vol. 5, No. 1, p. 65-80, 2010, 5, 65

\bibitem[Haffert et al. (2019)]{haffert19} Haffert, S.~Y., Bohn, A.~J., de Boaer,J., et al.\ 2019, Nature Astronomy, 329

\bibitem[Hauschildt et al.(1997)]{hauschildt97} Hauschildt, P. H., Baron, E.,  \& Allard, F.\ 1997, \apj, 483, 390

\bibitem[Helled \& Schubert(2009)]{helled2009} Helled, R., \& Schubert, G.\ 2009, \apj, 697, 1256

\bibitem[Helled, \& Bodenheimer(2011)]{helled11} Helled, R., \& Bodenheimer, P.\ 2011, \icarus, 211, 939


\bibitem[Hinkley et al.(2013)]{Hinkley13} Hinkley, S., Pueyo, L., Faherty, J.~K., et al.\ 2013, \apj, 779, 153 

\bibitem[Hoeijmakers et al.(2018)]{Hoeijmakers2018} Hoeijmakers, H.~J., Ehrenreich, D., Heng, K., et al.\ 2018, \nat, 560, 453

\bibitem[Jones et al.(2016)]{jones16} Jones, J., White, R.~J., Quinn, S., et al.\ 2016, \apjl, 822, L3

\bibitem[Kaeufl et al.(2004)]{kaeufl04} Kaeufl, H.-U., Ballester, P., Biereichel, P., et al.\ 2004, \procspie, 1218

\bibitem[Kesseli et al.(2019)]{kesseli19} Kesseli, A.~Y., Kirkpatrick, J.~D., Fajardo-Acosta, S.~B., et al.\ 2019, \aj, 157, 63


\bibitem[Karkoschka \& Tomasko(2010)]{karkoschka10} Karkoschka, E., \& Tomasko, M.~G.\ 2010, \icarus, 205, 674

\bibitem[Kirkpatrick et al.(2006)]{kirkpatrick06} Kirkpatrick, J.~D., Barman, T.~S., Burgasser, A.~J., et al.\ 2006, \apj, 639, 1120

\bibitem[Konopacky et al.(2013)]{Konopacky13} Konopacky, Q.~M., Barman, T.~S., Macintosh, B.~A., \& Marois, C.\ 2013, Science, 339, 1398


\bibitem[Krabbe et al.(2004)]{Krabbe04} Krabbe, A., Gasaway, T., Song, I., et al.\ 2004, \procspie, 5492, 1403 

%\bibitem[Larkin et al.(2006)]{Larkin06} Larkin, J., Barczys, M., Krabbe, A., et al.\ 2006, \nar, 50, 362 

\bibitem[Larkin et al.(2006)]{Larkin06} Larkin, J., Barczys, M., Krabbe, A., et al.\ 2006, \procspie, 6269, 62691A

\bibitem[Li et al.(2020)]{li20} Li, C., Ingersoll, A., Bolton, S., et al.\ 2020, Nature Astronomy, doi:10.1038/s41550-020-1009-3

\bibitem[Lockhart et al.(2019)]{Lockhart19} Lockhart, K.E., Do, T., Larkin, J.E., et al.\ 2019, \aj, 157, 75 

\bibitem[Madhusudhan et al.(2011)]{Madhu11} Madhusudhan, N., Mousis, O., Johnson, T.~V., \& Lunine, J.~I.\ 2011, \apj, 743, 191

\bibitem[Madhusudhan (2019)]{madhu19} Madhusudhan, N.\ 2019, \araa, 57, 617

\bibitem[Meyer et al.(2012)]{meyer12} Meyer, L., Ghez, A.~M., Sch{\"o}del, R. , et al.\ 2012, Science, 338, 84

\bibitem[Molli{\`e} et al.(2020)]{molliere20} Molli{\`e}re, P., Stolker, T., Lacour, S., et al.\ 2020, \aap, in press (arXiv:2006.09394)


\bibitem[Morbidelli et al.(2014)]{morbidelli14} Morbidelli, A., Szul{\'a}gyi, J., Crida, A., et al.\ 2014, \icarus, 232, 266

\bibitem[Mordasini et al.(2016)]{mordasini16} Mordasini, C., van Boekel, R., Molli{\`e}re, P., et al.\ 2016, \apj, 832, 41

\bibitem[Mousis et al.(2009)]{Mousis09} Mousis, O., Marboeuf, U., Lunine, J.~I., et al.\ 2009, \apj, 696, 1348

\bibitem[Nielsen et al.(2019)]{Nielsen19} Nikolov, E.~L., De Rosa, R.~J., Macintosh, B.~A., et al.\ 2019, \aj, 158, 13


\bibitem[Nikolov et al.(2018)]{nikolov18} Nikolov, N., Sing, D.~K., Fortney, J.~J., et al.\ 2018, \nat, 557, 526

\bibitem[{\"O}berg et al.(2011)]{oberg2011} {\"O}berg, K.~I., Murray-Clay, R., \& Bergin, E.~A.\ 2011, \apjl, 743, L16

\bibitem[Kosmo O'Neil et al.(2019)]{Kosmo-Oneil19} O'Neil, K. Kosmo, Martinez, G.~D., Hees, A., et al.\ 2019, \aj, 158, 4

\bibitem[Oreshenko et al.(2020)]{oreshenko20} Oreshenko, M., Kitzmann, D., Marquez-Neila, P., et al.\ 2020, \aj, 159, 6


\bibitem[Owen et al.(1999)]{Owen99} Owen, T., Mahaffy, P., Niemann, H.~B., et al.\ 1999, Nature, 402, 269

\bibitem[Petit dit de la Roche et al.(2018)]{PetitditdelaRoche2018} Petit dit de la Roche, D.~J.~M., Hoeijmakers, H.~J., \& Snellen, I.~A.~G.\ 2018, \aap, 616, 146

\bibitem[Piso et al.(2015)]{piso15} Piso, A.-M.~A., {\"O}berg, K.~I., Birnstiel, T., et al.\ 2015, \apj, 815, 109

\bibitem[Royer et al.(2007)]{royer07} Royer, F., Zorec, J., \& G{\'o}mez, A.~E.\ 2007, \aap, 463, 671

\bibitem[Ruffio et al.(2019)]{ruffio19} Ruffio, J.-B., Macintosh, B., Konopacky, Q.~M., et al.\ 2019, \aj, 158, 200

\bibitem[Santos et al.(2017)]{santos17} Santos, N.~C., Adibekyan, V., Figueira, P., et al.\ 2017, \aap, 603, A30

\bibitem[Snellen et al.(2014)]{snellen14} Snellen, I.~A.~G., Brandl, B.~R., de Kok, R.~J., et al.\ 2014, \nat, 509, 63

\bibitem[Szul{\'a}gyi et al.(2014)]{szulagyi14} Szul{\'a}gyi, J., Morbidelli, A., Crida, A., et al.\ 2014, \apj, 782, 65

\bibitem[Takeda \& Honda(2016)]{takeda16} Takeda, Y., \& Honda, S.\ 2016, \pasj, 68, 32 

\bibitem[Teague et al.(2019)]{teague19} Teague, R., Bae, J., \& Bergin, E.~A.\ 2019, \nat, 574, 378

\bibitem[Theissen \& West(2014)]{theissen14} Teague, C.~A \& West, A.~A.\ 2014, \apj, 794, 146

\bibitem[Todorov et al.(2016)]{todorov16} Todorov, K.~O., Line, M.~R., Pineda, J.~E., et al.\ 2016, \apj, 823, 14 

\bibitem[Uyama et al.(2020)]{uyama20} Uyama, T., Currie, T., Hori, Y., et al.\ 2020, \aj, 159, 40

\bibitem[Vigan et al.(2020)]{vigan20} Vigan, A., Fontanive, C., Meyer, M., et al.\ 2020, arXiv:2007.06573

\bibitem[Visscher \& Fegley(2005)]{visscher05} Visscher, C., \& Fegley, B., Jr.\ 2005, \apj, 623, 1221 

\bibitem[Wang et al.(2018a)]{wang2018} Wang, J.~J., Graham, J.~R., Dawson, R.., et al.\ 2018a, \aj, 156, 192

\bibitem[Wang et al.(2018b)]{WangJi2018} Wang, J., Mawet, D., Fortney, J.~J., et al.\ 2018b, \aj, 156, 272

\bibitem[Witte et al.(2011)]{witte11} Witte, S., Helling, C., Barman, T., Heidrich, N., \&  Hauschildt, P.\ 2011, \aap, 529, A44

\bibitem[Wong et al.(2004)]{wong04} Wong, M.~H., Mahaffy, P.~R., Atreya, S.~K., Niemann, H.~B., \& Owen, T.~C.\ 2004, \icarus, 171, 153 

\bibitem[Wu et al.(2011)]{wu11} Wu, Y., Singh, H.~P., Prugniel, P., et al.\ 2011, \aap, 525, A71


\bibitem[Yurchenko \& Tennyson(2014)]{yurchenko14} Yurchenko, S.~N., \& Tennyson, J.\ 2014, \mnras, 440, 1649

\bibitem[Zahnle \& Marley(2014)]{Zahnle14} Zahnle, K.~J., \& Marley, M.~S.\ 2014, \apj, 797, 41

\bibitem[Zuckerman et al.(2011)]{Zuckerman11} Zuckerman, B., Rhee, J.~H., Song, I., \& Bessell, M.~S.\ 2011, \apj, 732, 61 

        
\end{thebibliography}
\end{document}